\documentclass[
reprint,
superscriptaddress,
 amsmath,amssymb,
 aps,
 pra,
]{revtex4-2}

\usepackage[T1]{fontenc}
\usepackage{wasysym}
\usepackage{xcolor}
\usepackage{amsmath}
\usepackage{physics}
\usepackage{caption}
\usepackage{graphicx}
\usepackage{svg}
\usepackage{dcolumn}
\usepackage{bm}
\usepackage{hyperref} 
\usepackage{cleveref} 
\usepackage{orcidlink}

\newcommand{\wep}{W}
\newcommand{\gf}{\mathcal{W}}
\newcommand{\black}{\newmoon}
\newcommand{\white}{\fullmoon}

\newcommand{\aqa}{$\langle aQa ^L\rangle $ Applied Quantum Algorithms, Universiteit Leiden}
\newcommand{\lorentz}{Instituut-Lorentz, Universiteit Leiden, Niels Bohrweg 2, 2333 CA Leiden, Netherlands}

\begin{document}

\title{Sector length distributions of recursively definable graph states through analytic combinatorics}

\author{Elo\"ic Vall\'ee\,\orcidlink{0009-0005-2513-152X}}
\email{vallee@lorentz.leidenuniv.nl}
\affiliation{\aqa}
\affiliation{\lorentz}

\author{Kenneth Goodenough\,\orcidlink{0000-0002-1761-0038}}
\affiliation{Naturwissenschaftlich-Technische Fakult{\"a}t, Universit{\"a}t Siegen, Walter-Flex-Stra{\ss}e 3, 57068 Siegen, Germany}
\affiliation{College of Information and Computer Science, University of Massachusetts Amherst, 140 Governors Dr, Amherst, Massachusetts 01003, USA}

\author{Paul E. Gunnells\,\orcidlink{0000-0002-2742-3025}}
\affiliation{Department of Mathematics and Statistics, University of Massachusetts Amherst, 710 N. Pleasant St., Amherst, Massachusetts 01003, USA}

\author{Tim Coopmans\,\orcidlink{0000-0002-9780-0949}}
\affiliation{QuTech, Delft University of Technology, Lorentzweg 1, 2628 CJ Delft, The Netherlands}
\affiliation{EEMCS, Delft University of Technology, Mekelweg 4, 2628 CD Delft, The Netherlands}

\author{Jordi Tura\,\orcidlink{0000-0002-6123-1422}}
\affiliation{\aqa}
\affiliation{\lorentz}

\date{\today}

\begin{abstract}
The sector length distribution or Shor-Laflamme distribution (SLD) of quantum states is governed by the $k$-body correlations amongst the different systems, and has been used to study entanglement and error correction. 
A succinct description of a quantum state's SLD can be obtained by representing it through the coefficients of an appropriate weight enumerator polynomial, yielding bounds on fidelity under depolarizing noise and on multipartite entanglement. 
However, such expressions quickly grow out of hand and are generally difficult to achieve analytically, reflecting the computational hardness of the SLD.
We sidestep this problem and, instead of a \emph{single} state's SLDs, encode a \emph{family} of quantum state's SLD as a generating function.
We then find closed-form expressions for a large class of graph states which we call `recursively definable' and which include many common graphs such as path graphs, cycle graphs, star graphs, grid graphs, and more.
As direct corollary, we obtain analytical expressions for such graph states' concentratable entanglement, bounds on their depolarizing fidelity, and a multipartite entanglement criterion. 
Our work opens up the use of generating functions and more generally analytic combinatorics to solve problems in quantum information theory.

\end{abstract}

\maketitle

\section{Introduction}

Understanding how quantum entanglement give rise to correlations remains a central problem in quantum information theory~\cite{horodecki_quantum_2009}, with far-reaching consequences for quantum computation, secure communication, and foundational studies of nonclassical phenomena. 
A full characterization of correlations typically demands knowledge of the many-body wavefunction, whose description scales exponentially with system size. 
To navigate this challenge, sector length distributions (SLDs) offer a tractable and physically meaningful compromise. 
They quantify the contribution of $k$-body correlations in a basis-independent manner, providing a coarse-grained yet insightful perspective on the structure of multipartite entanglement. 
Furthermore, SLDs connect naturally to the weight enumerator polynomial (WEP), a powerful invariant widely employed in the theory of quantum error correction.

Originally introduced as invariant quantities in the study of quantum error-correcting codes~\cite{gottesman_stabilizer_1997}, SLDs and WEPs have since been employed to establish a wide range of results, including proving the non-existence of certain classes of codes~\cite{shor_quantum_1997} and revealing structural connections between graph codes and entanglement-distillation protocols in noisy quantum devices~\cite{goodenough_near-term_2024}.
In addition, WEPs can be directly connected to physical observables through Bell-sampling measurements~\cite{montanaro_learning_2017}, which has been implemented using trapped ions~\cite{miller_experimental_2024}, highlighting their operational relevance beyond purely mathematical analysis.
Beyond error correction, SLDs have been applied to entanglement detection~\cite{klockl_characterizing_2015,miller_shor-laflamme_2023}, the derivation of monogamy inequalities~\cite{serrano-ensastiga_multiqubit_2025}, and even in proving the nonexistence of a maximally entangled state of seven qubits~\cite{huber_absolutely_2017}. 
These results have further motivated research on SLDs as a concise method for characterizing quantum states~\cite{wyderka_characterizing_2020}.

However, while encoding information in terms of SLDs can substantially reduce the complexity of such problems, the main challenge shifts to computing the SLD itself, which is generally nontrivial. 
Several approaches have been studied before; for example, the work in~\cite{cao_quantum_2024} presents a tensor-network-based method, while~\cite{miller_shor-laflamme_2023} introduced enumeration techniques for graph states.

In this work, we showcase how insights from the field of analytic combinatorics~\cite{stanley_enumerative_1997,flajolet_analytic_2009} can be framed in the context of WEPs in a natural way.
More precisely, for certain sequences of graph states $\left(\ket{\Gamma_0}, \ket{\Gamma_1},\ldots\right)=\left(\ket{\Gamma_r}\right)_{r\in\mathbb{N}}$ we find the associated \emph{generating function}, denoted by $\mathcal{W}(x, y, z)$. 
This generating function captures information about the graph states $\ket{\Gamma_r}$, since---by definition---the $r$'th coefficient in its power series around $z=0$ equals the WEP of the state. 

Using techniques from analytic combinatorics, generating functions also provide information about the WEP for large $r$. 
A direct application is the derivation of closed-form expressions for the concentratable entanglement for all states $\ket{\Gamma_r}$ in the family.
In addition, singularity analysis of generating functions can be used to find exponentially tight bounds on the fidelity of states $\ket{\Gamma_r}$ under uniform depolarizing noise $\lambda$. 
Similar ideas also allow us to apply the purity criterion for entanglement from~\cite{horodecki_quantum_2009}. 
Interestingly, we find a type of phase transition in this criterion; for natural sequences $\left(\Gamma_r\right)_{r\in \mathbb{N}}$, either (almost) all graph states $\ket{\Gamma_r}$ fail or pass the criterion, depending on whether $\lambda < \lambda_c$ or $\lambda > \lambda_c$. 
We compute the explicit value of $\lambda_c$ in terms of the singularities of the derivatives of the generating function.

Our methods apply to path graphs, cycle graphs, star graphs, and more generally to any sequence of graphs that we call \emph{recursively definable}. 
Such graph states naturally arise in experimental setups, see~\cite{thomas_efficient_2022,thomas_fusion_2024,aqua_atom-mediated_2025}, and we prove that their associated generating functions are necessarily rational functions.
We extend~\cite{miller_shor-laflamme_2023} by deriving some of its results within a unified framework and by establishing the SLDs for new families of graph states.

Taken together, these results suggest that analytic combinatorics offers a new combinatorial perspective on studying correlations and entanglement across a wide range of quantum systems, potentially enabling new methods for both theoretical analysis and practical applications.

\section{Background}

\subsection{Graph states}

Given an $n$-vertex graph $\Gamma = (V,E)$, the associated \emph{graph state} is defined as the $n$-qubit quantum state~\cite{hein_multiparty_2004}
\begin{equation}
    \ket{\Gamma} := \prod_{(i,j) \in E} \text{CZ}_{i,j} \ket{+}^{\otimes |V|},
\end{equation}
where $\text{CZ}_{i,j}$ denotes the controlled-$Z$ operation acting on qubits $i$ and $j$.  
Graph states constitute a central class of multipartite entangled states and have been extensively studied in the context of measurement-based quantum computation, quantum error correction, and quantum network theory~\cite{hein_entanglement_2006,briegel_measurement-based_2009,meignant_distributing_2019}.

In this manuscript, we will focus on \emph{sequences of graph states}, i.e.~$\left(\ket{\Gamma_r}\right)_{r\in\mathbb{N}}$.
A typical example of such a sequence is the \emph{path graph} family $P_n$, which consists of $n$ vertices connected linearly in a chain. 
In this example, the parameter $r$ corresponds to the size of the $r$'th graph, i.e.~$n = r$, but one can consider more complex examples where the size depends non-trivially on $r$, i.e.~the size $n$ of the graph is given by some function $n(r)$. 

Several other sequences of graphs are of particular interest. 
The \emph{cycle graph} $C_n$ is obtained by connecting the endpoints of a path graph to form a closed loop. 
The \emph{star graph} $S_n$ is composed of a central vertex connected to $n-1$ peripheral vertices. 
Note that the above families can be defined recursively, e.g.~the path graph $P_{n+1}$ on $n+1$ vertices can be naturally constructed from $P_{n}$. 
Our results apply more generally to such recursively defined families of graphs, see \Cref{subsec:general_method} for more details. 
Among these, we can find the $n \times k$ grid graphs and the Pusteblume graph that have been studied in previous work~\cite{miller_shor-laflamme_2023}.
The additional examples studied in this paper are the \emph{complete bipartite graph} and the \emph{joint-squares graph}.
Recursively defined graph families allow for more general constructions such as \emph{extrusions}, \emph{revolutions} and \emph{edge splitting} (see  \Cref{fig:examples_recursive_graphs}).

\subsection{Hamming weight}

Any quantum state $\rho$ of $n$ qubits can be represented as a linear combination of Pauli strings. 
Specifically, we can write
\begin{equation}
    \rho = \frac{1}{2^n} \sum_{P \in \mathcal{P}^n} \Tr(P \rho) \, P,
\end{equation}
where $\mathcal{P}^n := \{I, X, Y, Z\}^{\otimes n}$ denotes the set of all $n$-qubit phaseless Pauli strings. 
This set forms an orthogonal basis with respect to the Hilbert-Schmidt inner product, satisfying $\Tr(P Q) = 2^n \, \delta_{P,Q}$.

To characterize the structure of Pauli strings, we define the \emph{Hamming weight} of a string $P$ as the number of non-identity operators it contains:
\begin{equation}
    \mathrm{wt}(P) = \bigl|\{ i : P_i \neq I \} \bigr|,
\end{equation}
where $P = P_1 \otimes \dots \otimes P_n$. 

\subsection{Sector-length distribution}
The \emph{sector length} $A_k$ describes the amount of $k$-body correlations in an $n$-qubit state $\rho$~\cite{aschauer_local_2004}:
\begin{equation} \label{eq:sld_definition}
A_k(\rho) := \sum_{P \in \mathcal{P}^n_{k}} |\Tr(P\rho)|^2
\end{equation}
where $\mathcal{P}^n_{k} := \{P \in \mathcal{P}^n : \text{wt}(P) = k\}$ is the set of Pauli strings with Hamming weight $k$.
In other words, this corresponds to all Pauli operators $P$ acting nontrivially on exactly $k$ parties.

The concept of SLDs also extends to error correction, where they are known as Shor–Laflamme enumerators (or Shor-Laflamme distributions, conveniently keeping the same acronym, SLD)~\cite{shor_quantum_1997,gottesman_stabilizer_1997}.
For an $n$-qubit stabilizer code with stabilizer group $\mathcal{S} \subset \{\pm 1, \pm i\}\otimes \mathcal{P}^N$, the sector-length $A_k$ corresponds to the number of elements in $\mathcal{S}$ with Hamming weight $k$, i.e.
\begin{equation}
    |\{ P \in \mathcal{S} : \text{wt}(P) = k \}| .
\end{equation}
In the case of a 1-dimensional codespace (i.e.~the stabilizer code defines a single state), this definition is equivalent to \Cref{eq:sld_definition}. 
As an example, since the stabilizer group of a Bell pair is given by $\lbrace{II, XX, -YY, ZZ\rbrace}$ we find that $A_0(\rho) = 1$, $A_1(\rho) = 0$, $A_2(\rho) = 3$.

Beyond the basic constraints, numerous additional relations among sector lengths are known; for a detailed characterization we refer the reader to Ref.~\cite{shor_quantum_1997,gottesman_stabilizer_1997,wyderka_characterizing_2020}.  
Although SLDs are not sufficient to fully characterize entanglement---for example, the tensor product of two Bell pairs and a $\mathrm{GHZ}_4$ state have the same SLD---they can nevertheless reveal some information about the entanglement structure. 

For example, they can be used to study the fidelity under uniform depolarizing noise, see \Cref{sec:fidelity_under_depolarizing_noise}.
Since the quantities $|\Tr(P \rho)|^2$ can be efficiently estimated in experiment using the Bell sampling method~\cite{montanaro_learning_2017,wang_efficient_2025}, these properties can be established without performing full state tomography.

\subsection{Weight enumerator polynomial}

It is customary to introduce a quantity related to the SLD known as the \emph{weight enumerator polynomial} (WEP)~\cite{shor_quantum_1997,gottesman_stabilizer_1997}. 
For a system of $n$ physical qubits, we define the WEP as
\begin{equation} \label{eq:wep_definition}
\wep(x,y) = \sum_{k = 0}^n A_{k} \, x^{n-k} y^k.
\end{equation}
Taking the Bell pair once more as an example, we have that $\wep(x, y)=x^2+3y^2$.

The WEP plays a crucial role in coding theory. 
It allows one to characterize the error-detecting and error-correcting capabilities of a code, and derive bounds on code performance. 
Moreover, in the quantum context, the WEP offers critical insights into the propagation of various error types through the code or state, as well as the effectiveness of their mitigation by the selected encoding scheme~\cite{shor_quantum_1997, goodenough_bipartite_2024}.

\subsection{Generating functions for weight enumerator polynomials of graph states} \label{subsec:generating_function}

As will be explained, the computation of SLDs of graph states can be elegantly reformulated as a type of counting problem of (improper) colourings on the underlying graph~\cite{miller_shor-laflamme_2023}. 
Such counting problems can be tackled using the framework of analytic combinatorics~\cite{stanley_enumerative_1997,flajolet_analytic_2009}. 
This framework relies on the concept of \emph{generating functions} (GF), a mathematical formalism that encodes a sequence of numbers or polynomials as the coefficients of a \emph{formal} power series, where ``formal'' indicates that convergence is not considered.
Generating functions provide a compact and expressive means to tackle certain counting problems.
As such, generating functions have seen widespread applications in combinatorics, but also more recently in the understanding of noise in quantum networks~\cite{goodenough_noise_2025}.

For a sequence of graph states $\left(\ket{\Gamma_r}\right)_{r \in \mathbb{N}}$, we define the generating function encoding its SLD as
\begin{equation}\label{eq:gf_of_wep}
\begin{aligned}
    \gf(x,y,z) := \sum_{r=0}^{\infty} \wep_{r}(x,y) \, z^r \\= \sum_{r=0}^{\infty} \sum_{k=0}^{n} A^{(r)}_{k} x^{n-k} y^{k} z^r
\end{aligned}
\end{equation}
where $\wep_{r}(x,y)$ is the WEP associated to $\Gamma_r$ and $A^{(r)}_{k}$ denotes the $k$'th sector length of the $n$-qubit state associated to the graph $\Gamma_r$.
Note that the number of qubits $n$ depends on $r$---we omit this dependence to simplify the notation. 
We note that multiple forms of generating functions exist; in the present discussion, we focus exclusively on \emph{ordinary generating functions}. 

The above generating function compactly encodes the WEPs and SLDs for all possible states within a given family.
In other words, a single function $\mathcal{W}(x, y, z)$ captures the full hierarchy of WEPs and SLDs across system sizes, in the sense that the WEP and SLDs can be retrieved by computing the derivatives
\begin{align}
\wep_{r}(x,y) =&~\frac{1}{r!} \frac{\partial^{r}\gf}{\partial z^r}\bigg|_{z=0} ,\\
    A_{k}^{(r)} =&~\frac{1}{r!k!} \frac{\partial^{r+k} \gf}{\partial z^r \partial y^k} \bigg|_{x,y,z = 1,0,0} .
\end{align}

Importantly, the power of generating functions goes beyond computing the desired expressions via appropriate derivatives; they can also be used to understand the asymptotic behaviour of the WEP without calculating derivatives explicitly, which becomes infeasible for large $n$. 
This, in turn, enables the study of the asymptotic entanglement of the associated states without computing individual WEPs, and facilitates the determination of concentratable entanglement, as well as tight bounds on the fidelity and noise thresholds for entanglement in graph states, see \Cref{sec:conc_ent,sec:fidelity_under_depolarizing_noise,sec:entanglement_criterion}.

In the particular case of an $n$-qubit graph state $\ket{\Gamma}$, the computation of the SLD can be mapped to a graph-theoretical colour assignment problem~\cite{miller_shor-laflamme_2023}: \textit{$A_k$ is equal to the number of black-white colour assignments of $\Gamma$ for which exactly $n-k$ white vertices have an even number of black neighbours}.
For the sake of clarity and convenience, we say that a vertex is \emph{admissible} if it is white and has an even number of black neighbours. 
This allows us to restate the theorem as follows: 
\textit{$A_{n-k}$ is equal to the number of colourings with weight $k$}, where the weight of a colouring is the number of admissible vertices. 

As an example, we consider the graph consisting of two vertices and a single edge, i.e.~a Bell pair up to local rotations. 
It can be seen that the four colourings labelled by $\left(\white \white \right)$, $\left(\white \black \right)$, $\left(\black \white \right)$ and $\left(\black \black \right)$ have $2$, $0$, $0$ and $0$ admissible vertices, respectively. 
Unfolding the definition, we thus find $A_0 = 1, A_1 = 0, A_2 = 3$, hence we recover the weight enumerator polynomial $W(x,y) = x^2+3y^2$, as expected.

With the above definition of admissible vertices in a colouring in mind, it is often more intuitive to interpret the generating function (GF) of a sequence of graph states in combinatorial terms, i.e.~as a function that counts colour assignments with certain weight.
In this definition, the variable $x$ keeps track of the number of admissible vertices and $y$ of the non-admissible vertices.
In general, the terminology generating function denotes any function that encodes sequences of numbers in the coefficients of a polynomial series.
For example, the functions defined in \Cref{eq:wep_definition,eq:gf_of_wep,eq:gf_fidelity_under_noise,eq:gf_entanglement_criterion} are examples of GFs.

\section{Explicit generating functions for families of graph states}

In this section we give three examples of graph families and their GFs, namely path, star and cycle graphs. 
We find their GFs using the \emph{transfer matrix formalism}, which has seen applications in both combinatorics and statistical physics. 
We show how this formalism extends to so-called recursively defined graphs as well. 
In Appendix~\ref{app:comb_proofs_GF}, we show how the generating functions of path, star, and cycle graphs can be obtained more directly using combinatorial proofs based on the so-called \emph{symbolic method}: a unified algebraic approach for constructing generating functions that count complex combinatorial objects by systematically decomposing and combining simpler ones according to established rules~\cite{flajolet_analytic_2009}.

\subsection{Path graphs}

We begin by examining the family of path graphs.
In this case, the family is parametrized by the number of qubits, i.e.~$r = n$.
Let us compute the first terms of the generating function.

The first graph in the family corresponds to the graph with $n=0$ vertices, i.e.~the empty graph.
For convenience, we take the convention $A^{(0)}_0 = 1$, which implies
\begin{equation} \label{eq:wep_0_vertex}
    \wep_{0}(x,y) = 1.
\end{equation}

\begin{figure}
    \centering
    \includegraphics[width=0.7\linewidth]{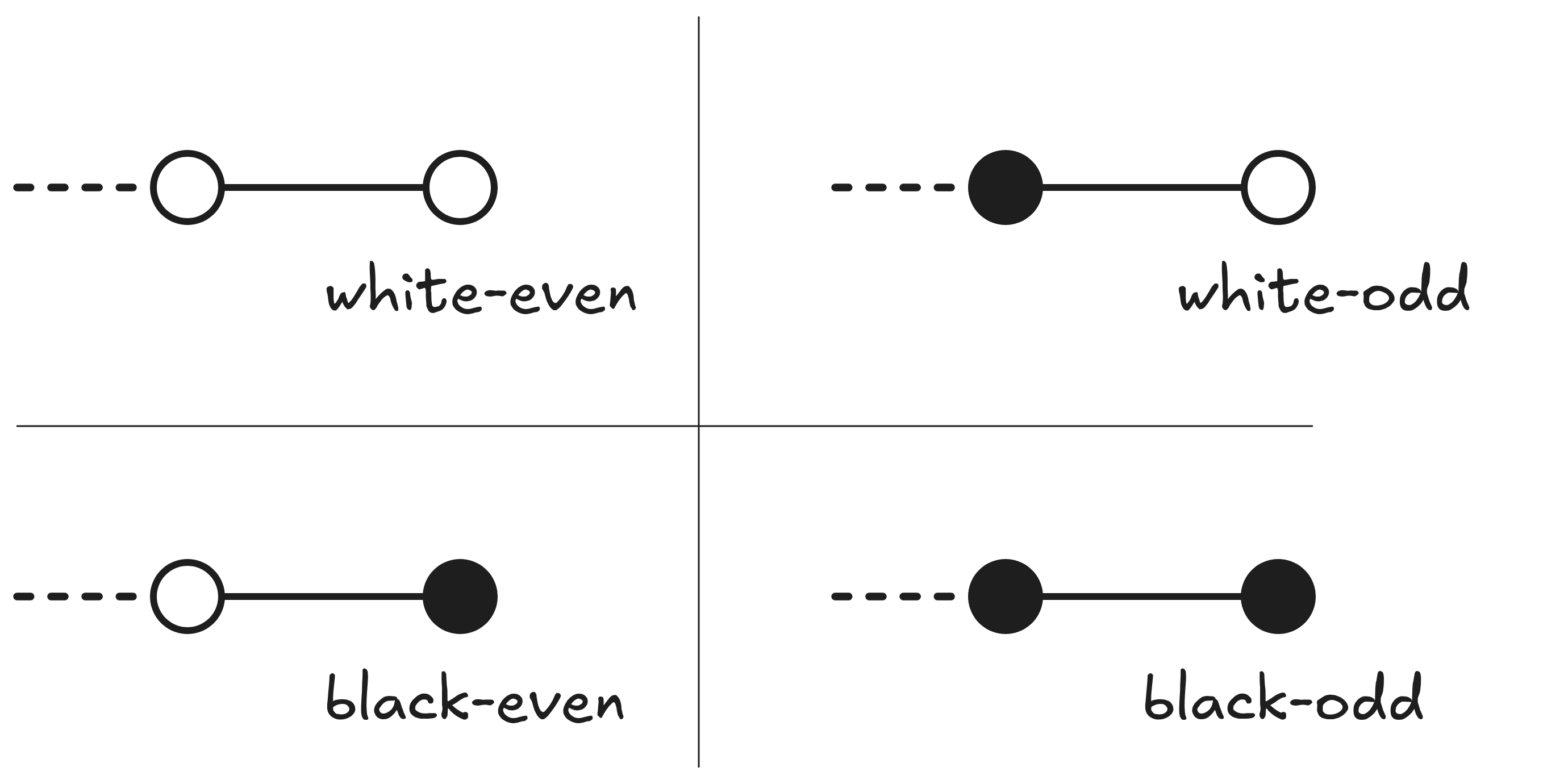}
    \caption{The generating function of the path graph can be sub-divided into four generating functions. 
    Each counting the number of colour assignment that end into one of the four possible states: white-even, white-odd, black-even, and black-odd. 
    The last vertex state is described by its colour and the parity of black neighbours.}
    \label{fig:possible_endings}
\end{figure}

For the single-vertex graph, there are two possible assignments: the vertex may be black or white.  
In the first case, there are no white vertices, giving weight zero; in the second, the vertex is white and has an even number (zero) of black neighbours, yielding weight one.  
Hence, $A^{(1)}_0 = A^{(1)}_1 = 1$ and 
\begin{equation} \label{eq:wep_1_vertex}
    \wep_{1}(x,y) = A^{(1)}_0 x + A^{(1)}_1 y = x + y.
\end{equation}
Similarly, we would find that $W_2(x, y)= x^2+3y^2$.
In principle, one could enumerate all possible colourings for higher $n$, but a more efficient strategy exploits the recursive structure of the path graph.  
Adding a black or white vertex to an existing colouring affects the total weight in a controlled manner, depending only on the colour and the parity of black neighbours of the preceding vertex. 
For all $n$-path graphs with $n \geq 1$, we can encode all possible colourings into a vector $\boldsymbol{v}^{(n)}$.
\Cref{fig:possible_endings} shows the four possible endings.
More precisely, we decompose the generating function into four terms, each counting assignments for which the last vertex is in a specific state.
The last vertex can be in four possible states: white-even, white-odd, black-even, and black-odd, which is described by its colour and the parity of black neighbours.

For example, consider the 1-path graph. 
The vertex has an even number of black neighbours (zero in this case), i.e.~there is no configuration in which the vertex will have an odd parity.
Therefore, the entries with odd parity are set to zero.
There is one configuration in state white-even.
In that case, the graph has exactly one admissible vertex, thus the GF for that entry is $x$.
There is one configuration in state black-even.
This corresponds to a non-admissible vertex; the associated GF is simply $y$.
The vector $\boldsymbol{v}^{(1)}$ of the 1-path graph is thus
\begin{equation} \label{eq:initial_vector_path_graph}
\boldsymbol{v}^{(1)} =
    \begin{pmatrix}
        x \\
        0 \\
        y \\
        0
    \end{pmatrix}.
\end{equation} 
To recover the GF, one must sum all components of the vector, this operation is denoted by $\| \cdot \|_{+}$.
For the 1-path graph we recover $\| \boldsymbol{v}^{(1)} \|_{+} = x + y = \wep_{1}(y)$, as expected.

The observation that the colouring weight can be updated by considering only the last vertex naturally leads to a recursive description.  
As we will show, the vector $\boldsymbol{v}^{(n+1)}$ can be obtained by multiplying $\boldsymbol{v}^{(n)}$ with a \emph{transfer matrix} $T$ encoding the effect of appending a new vertex (black or white) to each possible ending:
\begin{equation}
\boldsymbol{v}^{(n+1)} = T \boldsymbol{v}^{(n)} , \quad \forall n \geq 2
\end{equation}
The transfer matrix $T$ reflects the update rules for weights based on the vertex endings.
Each element $T_{ij}$ represents the contribution to the new ending $i$ from the previous ending $j$, with appropriate powers of $x$ and $y$.

As an illustration, the element $T_{1,1}$ describes the effect of appending a white vertex to a path graph ending with white-even.  
The resulting graph again ends in the state white-even, and the number of admissible vertices increases by one (due to an additional white vertex with an even number of black neighbours), which implies the introduction of a factor $x$.  
From this we find $T_{1,1} = x$. 
Another example is when the graph ends in the state white-even and we append a black vertex to it.
In that case, the new ending will be black and even (black-even).
The number of admissible vertices decreases by one (introducing a factor $x^{-1}$), while the number of non-admissible vertices increases by two, since the addition of a black vertex changes the parity of the last vertex from even to odd.
Therefore, the entry $T_{3,1}$ is given by $x^{-1}y^2$.
Performing this analysis for each possible case yields
\begin{equation}
    T = 
    \begin{pmatrix}
        x & x & 0 & 0 \\
        0 & 0 & y & y \\
        x^{-1}y^2 & x & 0 & 0 \\
        0 & 0 & y & y
    \end{pmatrix}.
\end{equation}

The generating function of the $n$-path graph is then obtained by summing all the entries of $\boldsymbol{v}^{(n)}$, i.e.,
\begin{equation}
\wep_{r}(x,y) = \left\| \boldsymbol{v}^{(r)} \right\|_{+} = \left\| T^{n-1} \boldsymbol{v}^{(1)} \right\|_{+} , \quad \forall n \geq 1
\end{equation}
Thus, the generating function of the path graph family is given by
\begin{equation}
\begin{aligned}
    \gf(x,y,z) = \sum_{r=0}^{\infty}  \wep_{r}(x,y)z^r = 1 + z \left\| \sum_{r = 1}^{\infty} (zT)^{r-1}\boldsymbol{v}^{(1)} \right\|_{+} ,\nonumber 
\end{aligned}
\end{equation}
where we used the convention in \Cref{eq:wep_0_vertex}.
Using the geometric-series identity for matrices, $\sum_{r=0}^{\infty} (zT)^r = (\mathbb{I}-zT)^{-1}$, we find
\begin{equation} \label{eq:gf_path_graph_family}
\begin{aligned}
    \gf(x,y,z) &= 1 + z \left\| (\mathbb{I}-zT)^{-1} \boldsymbol{v}^{(1)} \right\|_{+} \\
    &= \frac{1 - 2  (x - y) y z^2}{1 - z (x + y) (1 - (x - y) y z^2)}.
\end{aligned}
\end{equation}
The inverse matrix is computed by the adjugate–determinant formula for inverses, i.e.~$M^{-1} = \mathrm{adj}(M)/\det(M)$, where $\mathrm{adj}(M)$ denotes the adjugate of $M$~\cite{hohn2002elementary}.
Since both the adjugate and determinant are polynomial expressions in $x, y, z$, $W(x, y,z)$ is a rational function.

Path graphs are not the only family of graphs amenable to the above transfer matrix approach. 
More generally, we will see in \Cref{subsec:general_method} that there exist a broader class of graph families that can be tackled with a transfer matrix approach, and does have rational generating functions.

We close this section by noting that $\sum_{r=0}^{\infty} (zT)^r = (\mathbb{I}-zT)^{-1}$ might not converge (depending on the value of $z$). 
This is not an issue here, since generating functions are treated as formal power series and are therefore well defined for algebraic manipulation even without convergence~\cite{flajolet_analytic_2009}.

\subsection{Star graphs/GHZ states}

Star graph states are, up to local rotations, GHZ states. 
As before, the star graph family is also indexed by the number of vertices, i.e.~$r=n$.
Taking the same convention as before, the WEP of the graphs $n=0,1$ are given by \Cref{eq:wep_0_vertex,eq:wep_1_vertex}. 

For the transfer matrix approach, we keep track of the colour of the central vertex and the parity of its black neighbours.  
The colour of the central vertex is either white or black and its parity is even or odd.  
Hence, the basis vectors correspond to the states 
$\{\text{white-even}, \text{white-odd}, \text{black-even}, \text{black-odd}\}$.
In this basis, the initial vector encoding the one-vertex graph is the same as the initial vector of the path graph family given in \Cref{eq:initial_vector_path_graph}.

The transfer matrix is slightly different and takes the form
\begin{equation}
    T =
    \begin{pmatrix}
        x & x & 0 & 0 \\
        x^{-1}y^2 & x & 0 & 0 \\
        0 & 0 & y & y \\
        0 & 0 & y & y
    \end{pmatrix}
    =
    \begin{pmatrix}
    T_1 & 0 \\
    0 & T_2
\end{pmatrix},
\end{equation}
where we decomposed $T$ into $2 \times 2$ blocks:
\begin{equation}
T_1 = 
\begin{pmatrix}
    x & x \\
    x^{-1}y^2 & x
\end{pmatrix},
\qquad
T_2 = 
\begin{pmatrix}
    y & y \\
    y & y
\end{pmatrix}.
\end{equation}

The off-diagonal blocks are zero because they would correspond to transitions in which the colour of the central vertex changes.  
Since the colour of the central vertex is fixed once chosen, such transitions never occur.

The block $T_2$ corresponds to the case where the central vertex is black.  
All entries in this block are equal to $y$ because, regardless of the colour of the newly added vertex, the number of admissible vertices remains unchanged. 
In other words, adding a vertex will only increase the number of non-admissible vertices, which induces a factor $y$.

Conversely, the block $T_1$ corresponds to the case where the central vertex is white.  
In this case, one can verify that adding a new vertex increases the number of admissible vertices by one, i.e.~multiply the GFs by $x$, except when the central vertex has an even number of black neighbours, and a black vertex is added.  
In that specific configuration, the central vertex becomes surrounded by an odd number of black neighbours and thus is not admissible anymore.
Consequently, the number of admissible vertices decreases by one and the number of non-admissible vertices increases by two, i.e.~multiply the GFs by $x^{-1}y^2$.

The corresponding total generating function can be expressed as
\begin{equation} \label{eq:gf_star_graph_family}
\begin{aligned}
    \gf(x,y,z) 
     = \sum_{r=0}^{\infty} \wep_{r}(x,y) z^r
      = 1 + z \left \| (\mathbb{I} - zT)^{-1} \boldsymbol{v}^{(1)} \right\|_+ \\
     = \frac{yz}{1-2yz} + \tfrac{1}{2}\left(\frac{1}{1-(x+y)z}+\frac{1}{1-(x-y)z}\right)\ .
\end{aligned}
\end{equation}

\subsection{Cycle graphs}

Once again, the cycle graph family is parametrized by the number of vertices, i.e.~$r=n$.
In the literature, it is usually defined for $n \geq 3$.
For convenience, we define the $r=0$ (resp.~$r=1$) graph as the empty (resp.~one-vertex) graph and their WEPs are given by \Cref{eq:wep_0_vertex,eq:wep_1_vertex}.
Moreover, we define the $r=2$ graph as the graph of two isolated vertices.
The enumeration of the four possible colourings give the following WEP: 
\begin{equation} \label{eq:wep_2_vertex}
    \wep_{2}(y) = x^2 + 2xy + y^2 \; .
\end{equation}
This choice of the $r=2$ graph was made a posteriori to yield a GF of the cleanest form.

We now sketch how the transfer matrix approach can be extended to the family of cycle graphs, see Appendix~\ref{app:cycle_graph} for more details. 
The key idea is that, when a new vertex is added to complete the cycle, the update on the number of admissible/non-admissible vertices depends only on the first and last vertex. 
To apply the recursive procedure, we have to introduce a slight modification: instead of the parity of black neighbours, we will count the \emph{external parity}.
For an $n$-cycle graph, the external parity corresponds to the parity of black neighbours without counting the first and last vertices.
In other words, the external parity of the first vertex depends only on the colour of the second vertex.
Similarly, the external parity of the last vertex depends only on the colour of the $(n-1)$-th vertex.

We now let the transfer matrix keep track of the external parity as well. 
As we show in Appendix~\ref{app:cycle_graph}, a similar approach as with the previous two GFs can then be used to find the cycle graph GF. 
We find it is given by
\begin{equation} \label{eq:gf_cycle_graph_family}
\begin{aligned}
    \gf(x,y,z) & =  \sum_{r=0}^{2} \wep_{r}(x,y) z^r + z^3 \left\| (\mathbb{I} - zT)^{-1} \boldsymbol{v}^{(3)} \right\|_+ \\
    & = \frac{1 - 2 (x - y) (x + y) y z^3}{1 - z (x + y) (1 - (x - y) y z^2)}.
\end{aligned}
\end{equation}
The compactness of the above GF should be compared with the expression found for the individual $A_k^{(r)}$ in Eq.~21 of~\cite{miller_shor-laflamme_2023}.

\subsection{Transfer matrix approach: general method} \label{subsec:general_method}

Here we sketch the systematic approach that encompasses the path, cycle, star graphs, and extends to all sequences of graphs that are \emph{recursively definable}, which we define shortly. 
For a more detailed description, see Appendix~\ref{app:transfer_matrix}.

\begin{figure}
    \centering
    \includegraphics[width=0.9\linewidth]{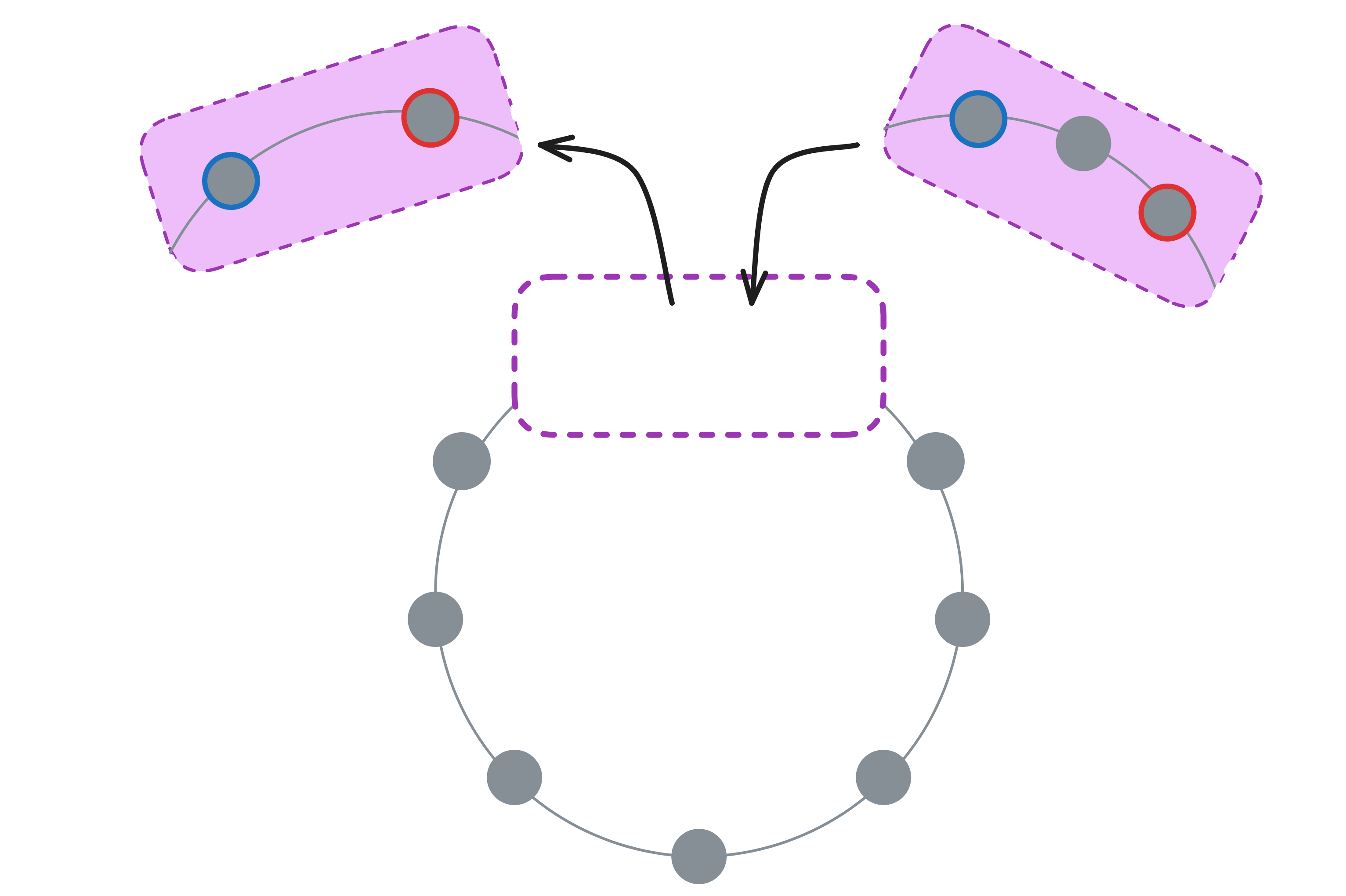}
    \caption{``Cut-and-glue'' construction. The cycle graph family can be constructed iteratively by removing the subgraph induced by the first and last vertices and replacing it with a path graph on three vertices. The newly introduced vertices are then connected to the existing ones according to a well-defined replacement map. In the figure above, this mapping is pictured by red and blue colors.}
    \label{fig:recursive_construction_cycle_graph}
\end{figure}

Intuitively, transfer matrices provide a natural formalism to describe \emph{cut-and-glue} operations. 
For example, the cycle graph family can be constructed iteratively by removing the subgraph generated by the first and the last node and replacing it with a path graph of three vertices. 
It is crucial, however, how these vertices are reattached to the remaining graph: connecting them in an arbitrary manner may produce a graph different from a cycle. 
It is therefore necessary to specify a \emph{replacement map} $\phi$, indicating which new vertex replaces the removed one.
The recursive construction of the cycle graph family can be seen in \Cref{fig:recursive_construction_cycle_graph}. 

A bit more formally, recursively defined graphs are defined as follows.
Given a graph $G$, we perform two operations: (i) select a set of vertices inducing a subgraph $H$; (ii) \emph{cut} the subgraph $H$ from $G$ and \emph{glue} a graph $J$ in its place, according to a pre-defined injective map $\phi$ from the vertices of $H$ to those of $J$. 
We show in Appendix~\ref{app:transfer_matrix} how to construct the transfer matrix associated to the above-mentioned cut-and-glue operations.

Several remarks are in order. 
For the above construction to be valid, there must exist an injective map $\phi$ from the vertices of $H$ to the vertices of $J$ (in particular, $|J| \geq |H|$). 
Second, this construction can be applied recursively by performing successive cut-and-glue operations on the newly added $J$. 
For this recursion to be well-defined, the graph $J$ must contain an induced subgraph $H'$ that is isomorphic to $H$, which we specify through another injective map $\phi'$ from the vertices of $H$ to the vertices of $J$.

We call sequences of graphs that can be defined in the above manner \emph{recursively definable}, and their associated generating functions can be naturally expressed within the transfer matrix formalism.
Path graphs, star graphs as well as cycle graphs are all three specific instances of recursively defined graphs.
The path graph can be constructed by recursively replacing the last node by two connected vertices. 
Analogously, the star graph is obtained by recursively replacing the central node by two connected vertices.

The cycle graph is built by recursively replacing two adjacent vertices by a 3-path graph.
Note that for the cycle graph family, the recursion starts at step $r=3$, the first three graphs being artificially fixed to the zero-, one- and two-disjoint-nodes graphs respectively.
For both the path graph and the star graph, the recursion begins at step $2$, whereas for the cycle graph, the recursion begins at step $3$.

As we show in Appendix~\ref{app:transfer_matrix}, the generating function for recursively defined graphs are necessarily rational functions.
In other words, after computing $(\mathbb{I}-zT)^{-1}$, the resulting expression of the GF is a ratio of polynomials, 
\begin{equation} \label{eq:GF_is_rational}
    \gf(x,y,z) = \frac{p(x,y,z)}{q(x,y,z)}\ .
\end{equation}
This follows from the fact that such sequences have associated transfer matrices, and that the associated generating functions involve the inverse of the matrix $\mathbb{I} - zT$, which---as before---is given by the adjugate-determinant formula. 
Matrix inverses are given by the ratio of their adjugate and determinant. 
Since the adjugate and determinant are polynomials, the claim follows.

\begin{figure}
    \centering
    \includegraphics[width=0.9\linewidth]{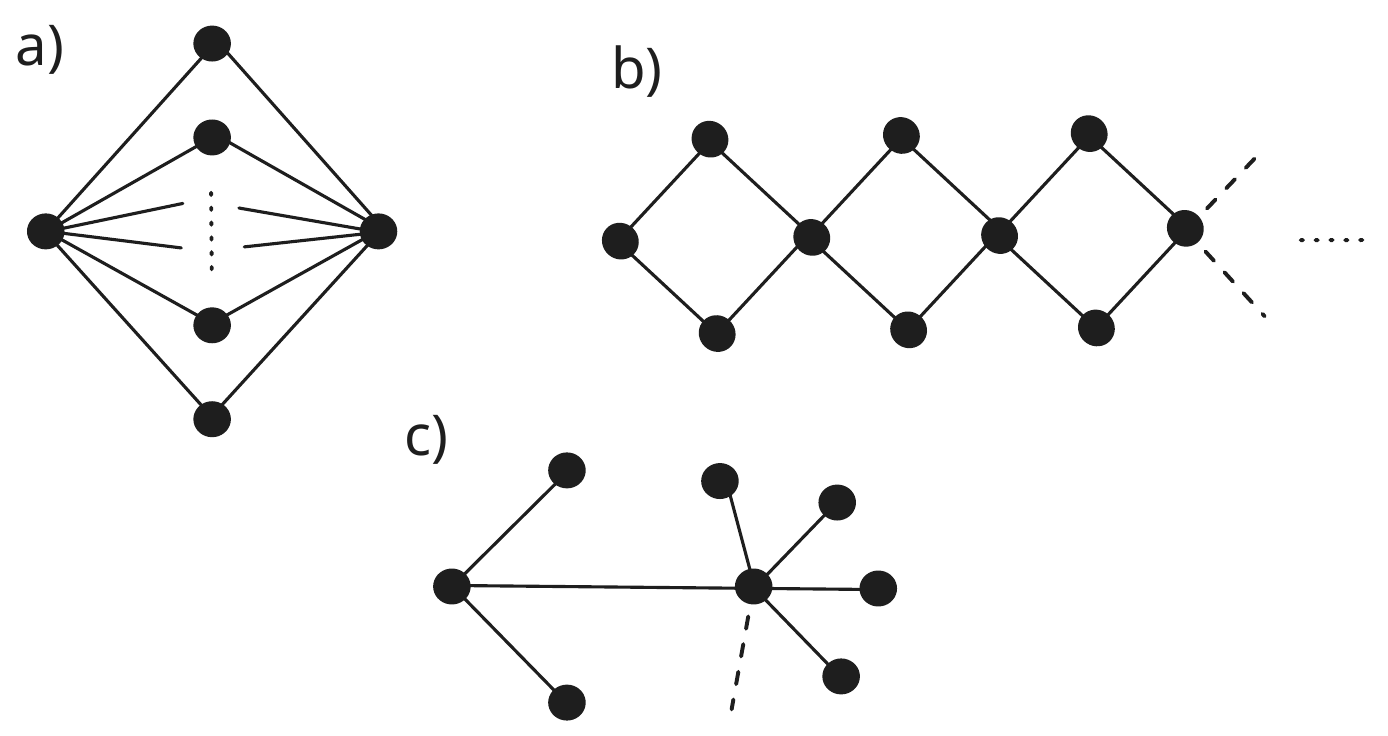}
    \caption{Examples of recursively defined graphs. 
    (a) complete bipartite graph. 
    The vertices can be divided into two disjoint sets such that no vertices within the same set are connected. 
    One of the sets is fixed to two vertices.
    (b) joint-squares graph. 
    Chain of 4-cycle graphs such that each consecutive pair shares a common vertex.
    (c) Pusteblume graph. Two leaves are attached to a leaf of a star graph. }
\label{fig:examples_recursive_graphs}
\end{figure}

The transfer matrix formalism allow us to express and study a number of new graph families.
In what follows, we present several examples of recursively defined graph families. 
This list is not intended to be exhaustive, but rather to illustrate the range of graph families that can be captured by the transfer-matrix approach.
The \emph{Pusteblume graph} is a tree consisting of a central vertex with three leaves, where one of these leaves is itself the center of a star graph. 
This graph is of particular interest because it has already been studied in the SLD literature (see, e.g.,~\cite{miller_shor-laflamme_2023}).
The \emph{complete bipartite graph} is a graph whose vertices can be divided into two disjoint sets such that every vertex in one set is connected to every vertex in the other set, and there are no edges within the same set.
For the sake of simplicity, we will consider that one of the sets only has two vertices.
The \emph{joint-squares graph} is formed by connecting a sequence of 4-cycle graphs such that each consecutive pair shares a common vertex.
These three graphs are shown in \Cref{fig:examples_recursive_graphs}.

Instead of focusing on specific families, we can also consider more general graph constructions that can be expressed recursively. 
An \emph{extrusion} corresponds to the Cartesian product of a graph (or subgraph) with a path graph, where each successive vertex in the path adds a copy of the original graph. 
A \emph{revolution} is similar to an extrusion, but the last copy of the graph is also connected to the first, corresponding to the Cartesian product with a cycle graph. 
The \emph{edge splitting} construction involves inserting a new node along an existing edge. 

The formal construction of generating functions for graph colourings is developed in detail in parallel work~\cite{goodenough_black-white_2026}.
Some of these constructions naturally arise in experimental setups~\cite{thomas_efficient_2022}. 
The generating functions of these graph families, as well as other families, can be found in \Cref{app:more_examples} and in the supporting code repository~\cite{vallee_code-recursive_2026}.

\section{Concentratable entanglement}\label{sec:conc_ent}
Concentratable entanglement is a multipartite entanglement measure~\cite{liu_generalized_2025}.
It is given by the average purity over all marginals of a pure state, and is thus given by
\begin{equation} \label{eq:concentratable_entanglement}
    C\left(\ket{\psi}\right) = 1 - \frac{1}{2^n} \sum_{\alpha \in 2^{\left[n\right]}} \Tr(\rho_{\alpha}^2)
\end{equation}
where $2^{\left[n\right]} = \{\emptyset, \{1\}, \{1,2\}, \dots, \{1,\dots,n\}\}$ is the power set of $n$ elements and $\rho_\alpha$ is the reduced density operator over the subsystem $\alpha$.
For convenience, we define $\overline{C}(\ket{\psi}) := 1-C(\ket{\psi})$.
Interestingly, the concentratable entanglement has an intuitive operational meaning~\cite{beckey_computable_2021}.
$\overline{C}(\ket{\psi})$ is equivalent to the probability of measuring the all-zeros bit string when performing a SWAP-test on each qubit between two copies of $\ket{\psi}$. 
As such, the concentratable entanglement can, in principle, be measured in an experiment.
Mathematically, we have
\begin{equation} \label{eq:conc_ent_def}
    \overline{C}(\ket{\psi}) = \Tr\left( \Pi^{\otimes n} \rho^{\otimes 2}\right)
\end{equation} 
where $\Pi = \frac{1}{2}(II+\operatorname{SWAP})$ is the measurement operator associated to the outcome 0 (also known as the symmetric projector)~\cite{buhrman_quantum_2001}. 
Here, each $\Pi$ acts on the same qubit $i$ from two copies of $\rho$. 
The symmetric projector in the Pauli basis is
\begin{equation}
    \Pi = \frac{3}{4}I I + \frac{1}{4}\left(XX + YY + ZZ\right) \ .
\end{equation}
The $n$-fold tensor product can then be expanded as
\begin{equation} \label{eq:n_fold_symmetric_projector}
   \begin{aligned}
       \Pi^{\otimes n} = \left(\frac{3}{4}I I + \frac{1}{4}\left(XX + YY + ZZ\right) \right)^{\otimes n}\\
= \sum_{P\in \mathcal{P}^n}\left(\frac{3}{4}\right)^{n-\mathrm{wt}(P)}\left(\frac{1}{4}\right)^{\mathrm{wt}(P)} P\otimes P \ .
   \end{aligned} 
\end{equation}
where $\mathcal{P}^n$ corresponds to the set of all $n$-qubit Pauli strings.
Moreover, in the Pauli basis, a state reads
\begin{equation}
    \rho = \frac{1}{2^n} \sum_{P \in \mathcal{P}^n} \Tr(P\rho) P \ ,
\end{equation}
and two copies of a state take the form
\begin{equation} \label{eq:copies_of_state_in_pauli_basis}
    \rho^{\otimes 2} = \frac{1}{2^{2n}} \sum_{P,Q \in \mathcal{P}^n} \Tr(P\rho)\Tr(Q\rho) P \otimes Q \ .
\end{equation}
Combining \Cref{eq:n_fold_symmetric_projector,eq:copies_of_state_in_pauli_basis} and the orthonormality of the Pauli matrices under the inner product $\langle U, V\rangle = \Tr(U^{\dagger} V)$, we get
\begin{equation}
\begin{aligned}
    \Tr\left(\Pi^{\otimes n}\rho^{\otimes 2}\right) & = \sum_{P \in \mathcal{P}^n} \left(\tfrac{3}{4}\right)^{n-\mathrm{wt}(P)}\left(\tfrac{1}{4}\right)^{\mathrm{wt}(P)} |\Tr(P \rho)|^2 \\
    & = \sum_{k = 0}^{n} \left( \tfrac{3}{4} \right)^{n-k} \left( \tfrac{1}{4} \right)^{k} A_{k} = W\left( \tfrac{3}{4}, \tfrac{1}{4} \right) \ .
\end{aligned}
\end{equation}
where $W$ is defined in \Cref{eq:wep_definition}.
Essentially, the concentratable entanglement of stabilizer states is an evaluation of their weight enumerator polynomial: $1-C = \overline{C} = W(\frac{3}{4},\frac{1}{4})$.

We can now define the generating function that encodes the concentratable entanglement for a family of recursively defined graph $\{\ket{\Gamma_r}\}_r$:
\begin{equation}
\begin{aligned}
    \overline{\mathcal{C}}(z) := \sum_{r=0}^{\infty} W_r\left( \frac{3}{4},\frac{1}{4} \right) z^r = \mathcal{W} \left( \frac{3}{4},\frac{1}{4},z \right)
\end{aligned}
\end{equation}
where $W_r$ is the WEP associated to the state $\ket{\Gamma_r}$.
The coefficient associated to $z^r$ corresponds to $1-C(\ket{\Gamma_r}) = \overline{C} = W_r (3/4, 1/4)$.
Closed-form expressions can be obtained using Cauchy’s residue theorem, which in particular expresses a rational function in terms of its singularities, see \Cref{app:asymptotic_behaviour_rational_functions}.
The generating functions appearing here are univariate and rational, and therefore possess only finitely many singularities. 
Although the locations of these singularities may be algebraically involved---typically given as roots of high-degree polynomials---this does not represent a fundamental obstruction. 

For graphs with additional structure, the singularity structure is often particularly simple, leading to correspondingly simple closed-form results.
We treat the path graph, star graph, and cycle graph as examples below.

\paragraph{Path graph.} The GF for the concentratable entanglement is obtained by making the replacement $x\mapsto \frac{3}{4}, y\mapsto \frac{1}{4}$ in \Cref{eq:gf_path_graph_family}
\begin{equation}
    \overline{\mathcal{C}}(z) = \frac{2z^2-8}{-z^3 + 8z - 8}
\end{equation}
which has singularities at $z_1 = -1+\sqrt{5}$ and $z_2 = -1-\sqrt{5}$, with multiplicity 1 each.
The concentratable entanglement of the $r$-th graph in the family corresponds to the coefficient associated to $z^r$ in the series expansion of $\overline{\mathcal{C}}(z)$.
It is given by 
\begin{equation}
\begin{aligned}
     \overline{C}_r & = \operatorname{Res}_{z=0} \left( \frac{\overline{\mathcal{C}}(z)}{z^{r+1}} \right) = - \sum_{i=1}^2 \operatorname{Res}_{z=z_i} \left( \frac{\overline{\mathcal{C}}(z)}{z^{r+1}} \right) \\
     & = \frac{5+\sqrt{5}}{5(-1+\sqrt{5})^{r+1}} + \frac{5-\sqrt{5}}{5(-1-\sqrt{5})^{r+1}}
\end{aligned}
\end{equation}
The second equality is a direct consequence of Cauchy’s residue theorem.
We note that the residues can be easily computed exactly using a compute algebra system, especially for rational functions.

\paragraph{Star graph.} After substituting $x\mapsto \frac{3}{4}, y\mapsto \frac{1}{4}$ in \Cref{eq:gf_star_graph_family}, we get $\overline{\mathcal{C}} = \frac{-z^2 - 2z + 4}{2z^2 - 6z + 4}$.
Singularities are $z_1 = 1$ and $z_2 = 2$, both have multiplicity 1.
Using Cauchy's residue theorem, we get
\begin{equation}
    \begin{aligned}
        \overline{C}_r = \frac{1}{2} + \frac{1}{2^r} \ .
    \end{aligned}
\end{equation}
This result is in perfect agreement with the value derived in earlier work~\cite{cullen_calculating_2022}.

\paragraph{Cycle graph.} 
The replacement $x\mapsto \frac{3}{4}, y\mapsto \frac{1}{4}$ in  \Cref{eq:gf_cycle_graph_family} yields $\overline{\mathcal{C}}(z) = \frac{-2z^3 + 8}{z^3 -8z +8}$, which has singularities at $z_1 = -1+\sqrt{5}$, $z_2 = -1-\sqrt{5}$ and $z_3 = 2$, with multiplicity 1.
Using Cauchy's residue theorem, we get 
\begin{equation}
    \overline{C}_r = 2^{-r} + \left( -1-\sqrt{5} \right)^{-r} + \left( -1+\sqrt{5} \right)^{-r} \ .
\end{equation}
Interestingly, the above expression can be rewritten as
\begin{equation}
    \begin{aligned}
        \overline{C}_r & = \frac{\phi^r+\left(\phi'\right)^r+1}{2^r} = \frac{L_r+1}{2^r} ,
    \end{aligned}
\end{equation}
where $\phi=1.618\ldots, \phi'=-0.618\ldots$ are the golden ratio and its conjugate, and $L_r$ is the $r$'th Lucas number, defined by the recursion $L_r=L_{r-1}+L_{r-2}$ and initial conditions $L_0=2, L_1=1$.

\section{Fidelity under depolarizing noise} \label{sec:fidelity_under_depolarizing_noise}

Understanding the effect of depolarizing noise on quantum states is crucial for assessing the robustness of quantum information processing. 
In particular, for large quantum systems, even small amounts of depolarizing noise can lead to significant degradation of quantum states, making the study of fidelity under such noise essential for quantum error correction and benchmarking protocols. 
While the fidelity of a state under uniform depolarizing noise can be calculated knowing the WEP (as we review here shortly), this calculation is still prohibitive for large quantum systems~\cite{shor_quantum_1997}. 
Instead, we bypass this problem by finding exponentially tight bounds on the fidelity, which are easy to compute in terms of the generating function.

For our noise model we assume depolarizing noise, a widely-used noise model that assumes isotropic errors, which uniformly degrade quantum states toward the maximally mixed state. 
Mathematically, the single-qubit depolarizing channel is described as~\cite{bruss_quantum_2000}:
\begin{equation}
    \mathcal{E}^{(1)}(\rho) = \lambda \rho + (1-\lambda) \frac{\mathbb{I}}{2},
\end{equation}
where $\rho$ is the single-qubit density matrix. 
In the Pauli basis, this translates to
\begin{equation}
    \mathcal{E}^{(1)} \left( \frac{1}{2} \sum_{P \in \mathcal{P} } \Tr(P\rho) P \right) = \frac{1}{2} \sum_{P \in \mathcal{P} } \lambda^{\mathrm{wt}(P)} \Tr(P\rho) P
\end{equation}
where $\mathcal{P} = \{I,X,Y,Z\}$ is the set of Pauli operators and the identity.
In other words, the depolarizing channel in the Pauli basis acts as $I \mapsto I$, $X \mapsto \lambda X$, $Y \mapsto \lambda Y$ and $Z \mapsto \lambda Z$.  

For an $n$-qubit system, the depolarizing channel acts independently and uniformly on each qubit,
\begin{equation}
    \mathcal{E}^{(n)}(\rho) = \bigotimes_{i=1}^n \mathcal{E}^{(1)}(\rho)\ .
\end{equation}
Its action on a Pauli string $P$ is given by $P \mapsto \lambda^{\mathrm{wt}(P)} P$, where $\mathrm{wt}(P)$ is the Hamming weight of $P$. 
Expressed in the Pauli basis, the channel acts on $\rho$ as
\begin{equation}
    \mathcal{E}(\rho) = \frac{1}{2^n} \sum_{P \in \mathcal{P}^n} \Tr(P \rho) \lambda^{\mathrm{wt}(P)} P.
\end{equation}

The inner product of the original state with its noisy counterpart is
\begin{equation} \label{eq:depolarizing_noise_inner_product}
    \begin{aligned}
        \Tr(\mathcal{E}(\rho)\rho) &= \frac{1}{2^n} \sum_{P \in \mathcal{P}^n} |\Tr(P\rho)|^2 \lambda^{\mathrm{wt}(P)}\\
    &= \frac{1}{2^n} \sum_{k=0}^n A_k \lambda^k = W\left(\frac{1}{2},\frac{\lambda}{2} \right)\ .
    \end{aligned}
\end{equation} 
The first equality is obtained by expressing both states in the Pauli basis, the second equality is obtained by grouping terms by their Hamming weight and the last equality is simply \Cref{eq:wep_definition}.
For pure states, \Cref{eq:depolarizing_noise_inner_product} coincides with the fidelity of the state under depolarizing noise. 
Hence, the fidelity of a pure state after uniform depolarizing noise with parameter $\lambda$ can be directly obtained from its (normalized) WEP.

We now show how the GF formalism allows us to find the asymptotic behaviour of families of recursive graph states.
Suppose we have a family $\left(\ket{\Gamma_r}\right)_r$ that is recursively definable.
From \Cref{eq:depolarizing_noise_inner_product} we know that the fidelity of $\ket{\Gamma_r}$ under uniform depolarizing noise $\lambda$ is given by
\begin{equation}
    F_r(\lambda) = \frac{1}{2^{n}} \sum_{k=0}^{n} A_k^{(r)} \lambda^k = \wep_{r}\left(\frac{1}{2},\frac{\lambda}{2} \right) \, .
\end{equation}
The sequence of fidelities $\{F_r(\lambda)\}_r$ can then be encoded into a generating function:
\begin{align} \label{eq:gf_fidelity_under_noise}
    \mathcal{F}(z;\lambda) :=& \sum_{r=0}^{\infty} F_r(\lambda) z^r\\
    = \gf\left( \tfrac{1}{2},\tfrac{\lambda}{2} , z\right) =&~ \frac{p(\tfrac{1}{2},\tfrac{\lambda}{2},z)}{q(\tfrac{1}{2},\tfrac{\lambda}{2},z)} \, .
\end{align}
where $p(x,y,z)$ and $q(x,y,z)$ are the numerator and denominator of $\gf(x,y,z)$ defined in \Cref{eq:GF_is_rational}.

Since the generating function $\mathcal{F}$ is rational, it is possible to extract the asymptotic behaviour of the coefficients $F_r(\lambda)$ as $r \rightarrow \infty$ (details are provided in \Cref{app:asymptotic_behaviour_rational_functions}).
For all the graph families we studied, we observed that the dominant singularity is unique and has multiplicity $m = 1$.
In this case, the asymptotic behaviour is approximately exponential: it is given by
\begin{equation}
    F_r(\lambda) \sim F_r^{\text{approx}}(\lambda) :=  - \frac{p(\tfrac{1}{2},\tfrac{\lambda}{2},z_*)}{\partial_z q(\tfrac{1}{2},\tfrac{\lambda}{2},z_*)}  z_*^{-r-1} \, ,
\end{equation}
where $\sim$ means that $\lim_{r \rightarrow \infty}F_r/F_r^{\text{approx}} = 1$, and where $z_*$ is the root of $q(\tfrac{1}{2},\tfrac{\lambda}{2},z)$ with the smallest modulus, i.e.~$z_* = \arg \min |z|$ s.t.~$q(\tfrac{1}{2},\tfrac{\lambda}{2},z) = 0$.
In particular, $z_*$ might depend on $\lambda$. 

\begin{figure}
    \centering
    \includegraphics[width=1\linewidth]{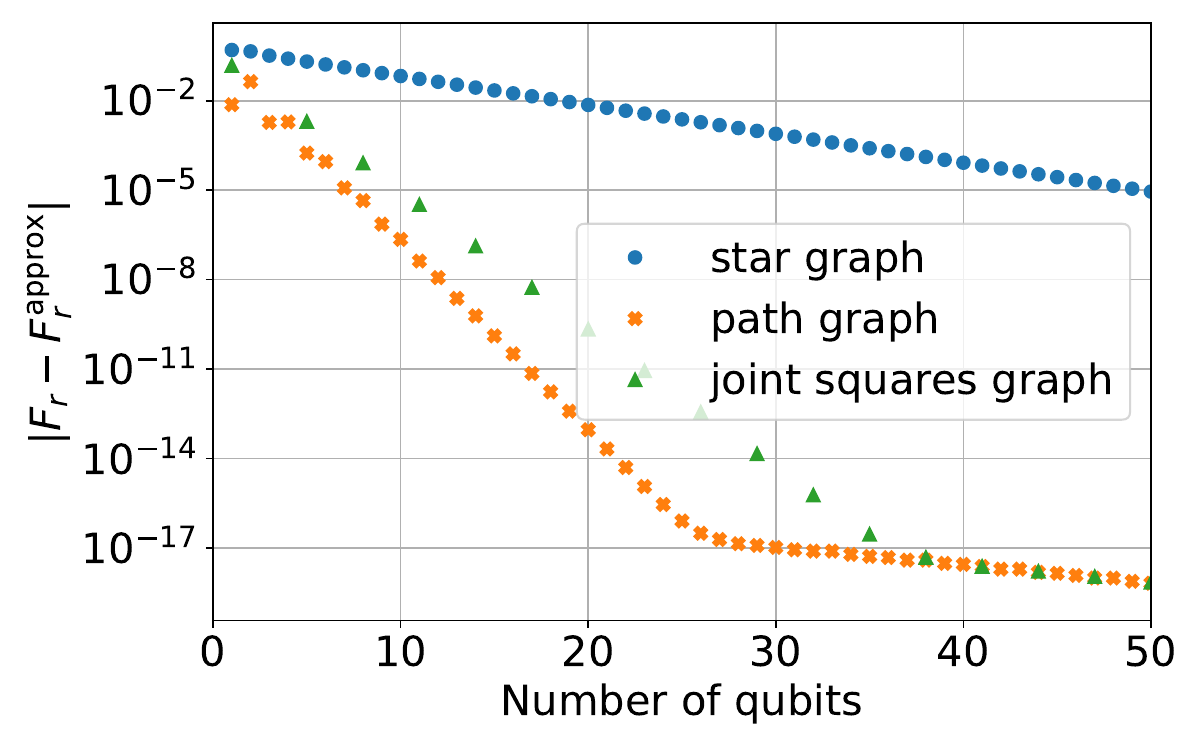}
    \caption{Difference between exact and approximated coefficients of the fidelity under depolarizing noise ($\lambda = 0.8$) as function of the number of qubits. 
    The approximation of the fidelity converges exponentially towards the exact value as the number of qubits increases.}
    \label{fig:convergence_fidelity}
\end{figure}

\Cref{fig:convergence_fidelity} shows the difference between the exact and approximated coefficients of the fidelity under depolarizing noise ($\lambda = 0.8$) as function of the number of qubits: $\Delta = |F_r - F_r^{\text{approx}}|$. 
The approximation of the coefficients converges exponentially fast towards the exact value as $r$ increases.

We note that better approximations can be found by considering more roots of $q(\frac{1}{2}, \frac{\lambda}{2}, z_*)$~\cite{flajolet_analytic_2009}. 
Furthermore,  we assumed in the above that the sequences of graph states under consideration were recursively definable, such that their GF is rational. 
While the analysis of rational GFs is particularly simple, we note that analytic combinatorics also allows for the asymptotic study of GFs that are more complex~\cite{flajolet_analytic_2009}, such that our analysis can in principle be extended to more complex sequences of graph states.

\section{Entanglement criterion} \label{sec:entanglement_criterion}
Due to the inherent noise in quantum systems, it is important to understand in which noise regimes a given state is still entangled or not.
In~\cite{miller_shor-laflamme_2023}, the authors derive an entanglement criterion based on the purity criterion~\cite{horodecki_quantum_2009}. 
More specifically, they showed that a state $\rho$ undergoing depolarizing noise with parameter $\lambda$ is entangled if
\begin{equation}\label{eq:ent_cond}
     Q = \sum_{k=0}^n (n-2k) A_k(\rho) \lambda^{2k} < 0 \, .
\end{equation}
Equivalently, we can express the above quantity as a difference $Q = Q_1-Q_2$, with
\begin{equation}
    Q_1 = \sum_{k=0}^{n} (n-k) \lambda^{2k} A_k(\rho)  \, , \, \, Q_2 = \sum_{k=0}^{n} \lambda^{2k} k A_k(\rho) \ . 
\end{equation}
such that $Q_1-Q_2 < 0$ implies that $\rho$ is entangled.
We define the \emph{critical noise parameter} as the value of $\lambda$ at which a state $\ket{\Gamma}$ becomes entangled, 
\begin{equation}
    \lambda_c := \{\lambda|Q_{1}(\lambda)-Q_{2}(\lambda) = 0, \;0 \leq \lambda \leq 1\} .
\end{equation}
If $\lambda > \lambda_c$, the state is entangled. 
Note that the converse statement cannot be made, since the entanglement criterion is only a sufficient, but not necessary, condition. 
That is, if $\lambda < \lambda_c$, it is not known whether the state is entangled or separable.

For a sequence of recursively defined graph states $\left(\ket{\Gamma_r}\right)_r$, we define the following GFs
\begin{equation} \label{eq:gf_entanglement_criterion}
\begin{aligned}
    \mathcal{Q}_1(z;\lambda) & = \sum_{r=0}^{\infty} \underbrace{\sum_{k=0}^n \lambda^{2k} (n-k) A_k^{(r)}}_{Q_{1,r}(\lambda)} z^r= \frac{\partial \gf}{\partial x} (1,\lambda^{2},z)  \\ 
    \mathcal{Q}_2(z;\lambda) & = \sum_{r=0}^{\infty} \underbrace{\sum_{k=0}^n \lambda^{2k} k A_k^{(r)}}_{ Q_{2,r}(\lambda)} z^r = \lambda^2 \frac{\partial \gf}{\partial y} (1,\lambda^2,z)
\end{aligned}
\end{equation}
where $A_k^{(r)} := A_k(\ket{\Gamma_r})$. 
Additionally, the critical noise parameter naturally becomes a function of the graph index $r$: $\lambda_c(r)~\in~[0,1]$ is the solution to the equation $Q_{1,r}~-~Q_{2,r}~=~0$.
Since $\gf$ is rational, the two GFs $\mathcal{Q}_1$ and $\mathcal{Q}_2$ in \Cref{eq:gf_entanglement_criterion} are also rational, and we have
\begin{equation}
\begin{aligned}
    \mathcal{Q}_1(z;\lambda) & = \frac{(q \partial_x p - p \partial_x q)(1,\lambda^2,z)}{q^2(1,\lambda^2,z)} \\
    \mathcal{Q}_2(z;\lambda) & = \lambda^2 \frac{(q \partial_y p - p \partial_y q)(1,\lambda^2,z)}{q^2(1,\lambda^2,z)} \\
\end{aligned}
\end{equation}
where $p(x,y,z)$ and $q(x,y,z)$ are the numerator and denominator of $\gf(x,y,z)$ defined in \Cref{eq:GF_is_rational}.

Analogously to the previous section, we use the fact that recursively definable graph states have rational generating functions, such that we can approximate their coefficients $Q_{1,r}(\lambda)$ and $Q_{2,r}(\lambda)$ as
\begin{equation}
    \begin{aligned}
        Q_{1,r}(\lambda) & \sim (-1)^m \frac{m (q \partial_x p - p \partial_x q)(1,\lambda^2,z_*)}{(\partial_z^m q^2)(1,\lambda^2,z_*)} r^{m-1} z_*^{-r-m} \\
        Q_{2,r}(\lambda) & \sim (-1)^m \frac{m \lambda^2 (q \partial_y p - p \partial_y q)(1,\lambda^2,z_*)}{(\partial_z^m q^2)(1,\lambda^2,z_*)} r^{m-1} z_*^{-r-m}
    \end{aligned}
\end{equation}
where $z_*$ is the root of $q^2(1,\lambda^2,z)$ with the smallest modulus and $m$ is its multiplicity.
Note that $z_*$ depends on the noise parameter $\lambda$.
The method to compute these values is explicitly given in \Cref{app:asymptotic_behaviour_rational_functions}.

With the above in mind, \Cref{eq:ent_cond} leads to the following entanglement criterion. 
In the limit of large $r$, the entanglement criterion is well approximated by
\begin{equation}\label{eq:ent_criterion}
    1 > \frac{Q_{1,r}(\lambda)}{Q_{2,r}(\lambda)} = \frac{(q \partial_x p - p \partial_x q)(1,\lambda^2,z_*)}{\lambda^2 (q \partial_y p - p \partial_y q)(1,\lambda^2,z_*)} \; .
\end{equation}
In other words, when $r$ is sufficiently large (corresponding in fact to a large number of qubits), the state $\ket{\Gamma_r}$ is entangled if \Cref{eq:ent_criterion} is satisfied.

Surprisingly, the \emph{approximated criterion} does not depend on the graph index $r$. 
This implies the existence of an \emph{approximated critical noise parameter} $\lambda_c^{\text{approx}}$ such that, for a sufficiently large $r_0$ (i.e.~a large number of qubits), all the state $\{ \ket{\Gamma_r} \}_{r>r_0}$ will be entangled whenever $\lambda > \lambda_c^{\text{approx}}$.
This behaviour is characteristic of a \emph{phase transition}: all states with a large number of qubits satisfy the entanglement criterion whenever $\lambda > \lambda_c^{\text{approx}}$.
Similar phase-transition-like behaviour has been observed in previous work~\cite{numajiri_phase_2025}.

\begin{figure}
    \centering
    \includegraphics[width=1\linewidth]{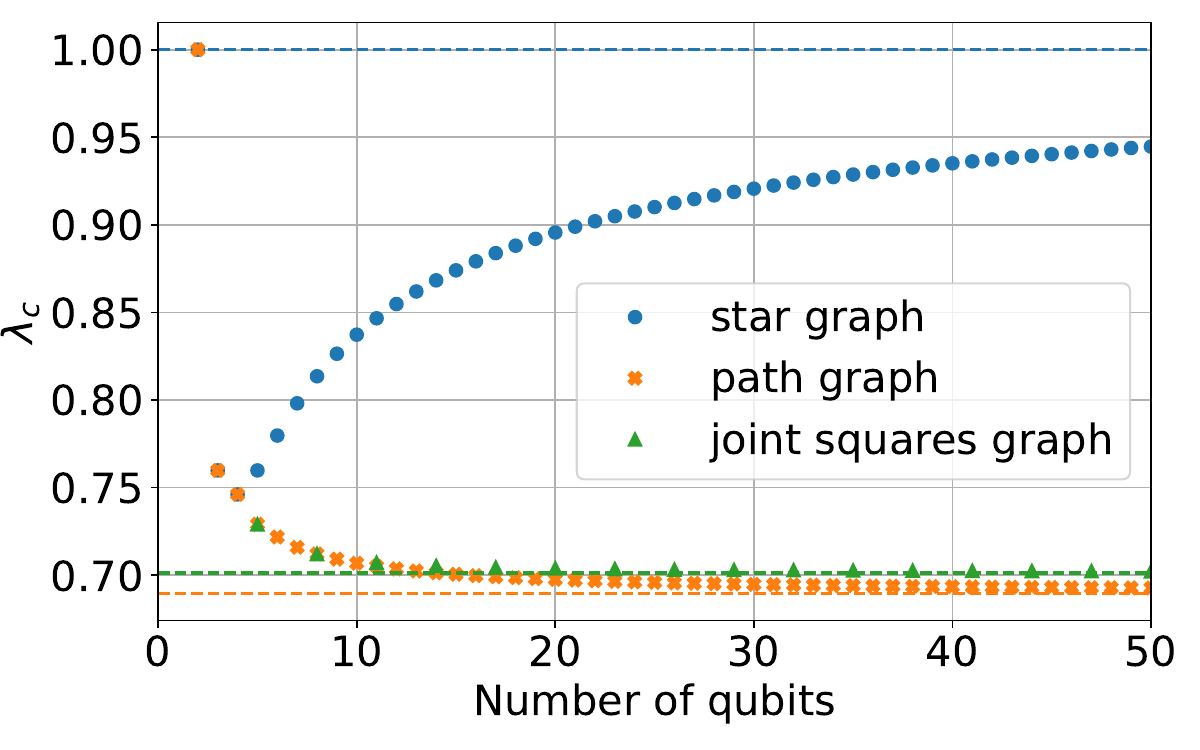}
    \caption{Critical noise parameter $\lambda_c$ as function of the number of qubits for different graph states families.
    When $\lambda > \lambda_c$, the state is entangled.
    In the asymptotic limit, $\lambda_c$ can be approximated using \Cref{eq:ent_criterion}.}
    \label{fig:critical_lambda}
\end{figure}

In \Cref{fig:critical_lambda}, we show the critical noise parameter, as well as the approximated noise parameter for the path graph, the star graph and the joint squares graph.
The behaviour of $\lambda_c(r)$ for cycle graphs is similar to that for path graphs.
For the complete bipartite graph, $\lambda_c(r)$ is similar to the one on the star graph.
For other graph families, we refer the reader to the online repository~\cite{vallee_code-recursive_2026}.

\section{Conclusion}
This work establishes a connection between quantum information and analytic combinatorics, providing a framework to compute and express weight enumerator polynomials for sequences of graph states by means of the generating function formalism.
We introduce recursively defined graph-state families, whose generating functions can be systematically obtained using the transfer matrix approach, enabling the construction and study of many structured families, including path graphs, star graphs, and cycle graphs.

Within this framework, we derive closed-form expressions for concentratable entanglement, accurately approximate fidelity under depolarizing noise in the many-qubit limit, and uncover a sharp phase-transition–like behaviour in entanglement: for a critical parameter $\lambda_c$, (almost) all states in the family are entangled when $\lambda > \lambda_c$.

Our results demonstrate that analytic combinatorics and generating functions provide a powerful and general tool for studying large-scale qubit systems. 
The approach opens the door to studying other generating functions, graph codes, and multi-parameter families, such as $n \times k$ grid states, offering a systematic path toward understanding scaling and robustness of entanglement in complex quantum systems.

\begin{acknowledgments}
J.T.~and E.V.~acknowledge the support received from the European Union's Horizon Europe research and innovation programme through the ERC StG FINE-TEA-SQUAD (Grant No.~101040729).
J.T. acknowledges the support received by the Dutch National Growth Fund
(NGF), as part of the Quantum Delta NL programme. 
K.G.~acknowledges support by the Alexander von Humboldt foundation.
T.C. acknowledges the support received through the NWO Veni programme (VI.Veni.232.381).

The views and opinions expressed here are solely those of the authors and do not necessarily reflect those of the funding institutions. 
None of the funding institutions can be held responsible for them.
\end{acknowledgments}

\bibliography{references-local,references-zotero} 
\newpage
\appendix

\section{Transfer matrix formalism for the cycle graph}\label{app:cycle_graph}
We detail here the derivation of the transfer matrix and the generating function of cycle graphs. 
Recall that for an $n$-cycle graph, the external parity corresponds to the parity of black neighbours without counting the first and last vertices.
That is, the external parity of the first vertex depends only on the colour of the second vertex, and the external parity of the last vertex depends only on the colour of the $(n-1)$-th vertex.

Since the update depends only on the first and last vertex, we divide the GF of an $n$-cycle graph in 16 GFs.
Each of them enumerates assignments whose first and last vertex are either black or white and have external parity either odd or even.
An example of a possible basis is $(c_1,c_n,p_1,p_n) \in \{0,1\}^4$, where $c_i$ (resp. $p_i$) denotes the colour (resp. external parity) of vertex $i$.
We will take the convention that $c_i = 0$ corresponds to the colour white and $p_i = 0$ corresponds to even parity.
The initial vector is found by explicitly computing all possible colourings of the 3-cycle graph and classifying them in one of the 16 categories according to the colour and external parity of the first and last vertex.
Explicitly, we get
\begin{equation}
    \boldsymbol{v}^{(3)} = (x^3,0,0,y^3,y^3,0,0,xy^2,y^3,0,0,xy^2,xy^2,0,0,1)^T \ . \nonumber
\end{equation}
The transfer matrix $T$ encodes the allowed transitions when adding a new vertex to the cycle. Each entry of $T$ updates the GF depending on the colour and the external parity of the first and last vertices.
For example, consider the transition from the $n$-cycle graph to the $(n+1)$-cycle graph.
The ending of the $n$-cycle graph is $(c_1^{(j)},c_n^{(j)},p_1^{(j)},p_n^{(j)})$, where $j \in \{1,\dots,16\}$.
Similarly, the ending of the $(n+1)$-cycle graph is denoted $(c_1^{(i)},c_{n+1}^{(i)},p_1^{(i)},p_{n+1}^{(i)})$, with $i \in \{1,\dots,16\}$.
The entry $T_{ij}$ will be non-zero only if the endings are compatible, i.e.~the colour and external parity of the first vertex must be the same for both graphs and the external parity of the $(n+1)$-th vertex must match the colour of the $n$-th node of the $n$-cycle graph.
This can be expressed explicitly as follows,
\begin{equation}
    T_{ij} = 
    \begin{cases}
        x^{\Delta} y^{1-\Delta}, & \text{if} \; c_1^{(i)} = c_1^{(j)}, p_1^{(i)} = p_1^{(j)}, p^{(i)}_{n+1} = c_{n}^{(j)} \\
        0, & \text{otherwise.}
    \end{cases}\nonumber 
\end{equation}
where $\Delta$ stands for the difference between the number of admissible vertices between the $n$-cycle graph and the $(n+1)$-cycle graph, and is explicitly given by
\begin{equation}
\begin{aligned}
    \Delta =  & +(1-c_1) (1 + p_1 + c_{n+1} \mod 2) \\
    & +(1-c_n) (1 + p_n + c_{n+1} \mod 2) \\
    & +(1-c_{n+1}) (1 + c_n + c_1 \mod 2) \\
    & -(1-c_1) (1 + p_1 + c_n \mod 2) \\
    & -(1-c_n) (1 + p_n + c_1 \mod 2) .
\end{aligned}
\end{equation}
Note that we simplified the notation, i.e.~$c_1 = c_1^{(i)} = c_1^{(j)}$, $p_1 = p_1^{(i)} = p_1^{(j)}$, $c_n = c_n^{(j)}$, $p_n = p_n^{(j)}$ and $c_{n+1} = c_{n+1}^{(i)}$.

As before, the terms corresponding to $n \geq 3$ can be expressed as a geometric series of the transfer matrix.
The full GF corresponds to
\begin{equation}
\begin{aligned}
    \gf(x,y,z) & =  \sum_{r=0}^{2} \wep_{r}(x,y) z^r + z^3 \left\| (\mathbb{I} - zT)^{-1} \boldsymbol{v}^{(3)} \right\|_+ \\
    & = \frac{1 - 2 (x - y) (x + y) y z^3}{1 - z (x + y) (1 - (x - y) y z^2)}.
\end{aligned}\nonumber
\end{equation}

\section{Transfer matrices for recursively definable graph states}\label{app:transfer_matrix}
Before showing how to construct the transfer matrix $T$ for a given sequence of recursively definable graphs, we recall some definitions to fix the notation.

A \emph{recursive graph} family can be defined in an iterative-constructive way.
Suppose we start from a graph G. 
A step goes as follows:
(i) identify a sub-graph $H$ of $G$.
(ii) replace $H$ by another graph $J$, according to a pre-defined injective map $\phi$ from the vertices of $H$ to those of $J$.
The recursion occurs when the start graph and the replacement graph are the same, i.e.~$G=J$.

For the needs of the construction, we also define the \emph{external parity} of a vertex in a sub-graph $H \subset G$. 
It is defined as the number (modulo 2) of black neighbours in $G \setminus H$.
Note that if the sub-graph actually corresponds to the full graph, i.e.~$H = G$, then the external parity of each node is fixed to zero (because there are zero black neighbours). 

\begin{figure}
    \centering
    \includegraphics[width=1\linewidth]{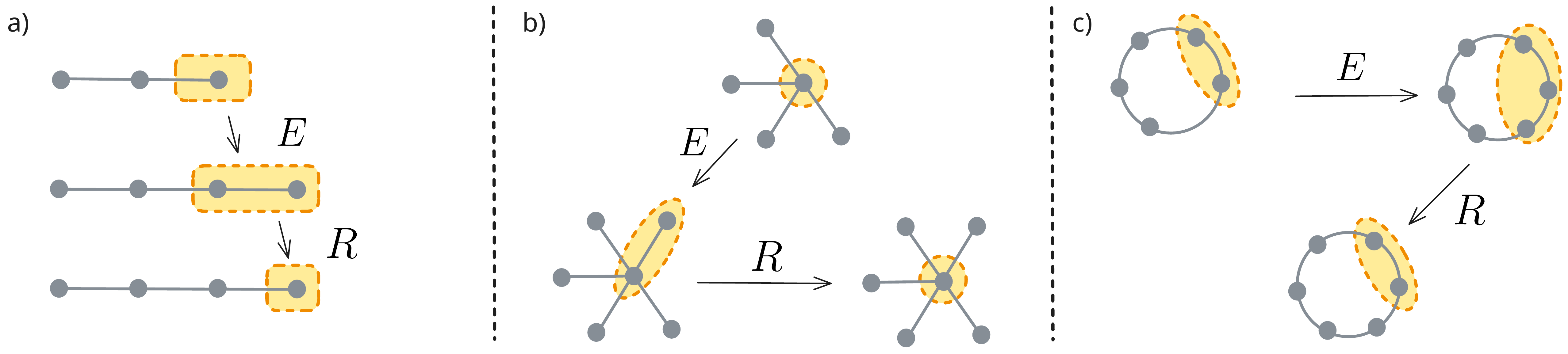}
    \caption{The path, star, and cycle graph families are typical instances of recursively defined graph.
    They can be generated through a common recursive construction procedure: an evolution step $E$ is first applied to a selected subset of vertices, followed by a restriction step $R$ that selects a new subset on which the procedure is applied recursively.}
    \label{fig:recursive_construction}
\end{figure}

We now show how to construct the transfer matrix $T$ for a given sequence of recursively definable graphs.
The construction proceeds by assigning to each node of every graph in the sequence two characteristics: a colour (black or white) and an external parity (even or odd).
Consequently, each node can occupy one of four states, $S = \{\text{we}, \text{wo}, \text{be}, \text{bo}\}$, and a graph $H$ has $4^{|H|}$ possible states. 
The weight of a graph $H$ in a given state $s \in S^{|H|}$ is defined as
\begin{equation}
    w(H,s) = \sum_{k \in V(H)} (1-c_k) \Big(1 + p_k + \sum_{l \in N_G(k)} c_l \bmod 2 \Big),\nonumber 
\end{equation}
where $V(H)$ denotes the vertices, $N_H(k)$ denotes the neighbours of vertex $k$, and
\begin{equation}
c_k = 
\begin{cases}
0, & k \text{ is white,} \\
1, & k \text{ is black,}
\end{cases}
\quad
p_k = 
\begin{cases}
0, & k \text{ is even,} \\
1, & k \text{ is odd.}
\end{cases}
\end{equation}
The values of $\{c_k\}_{k=1}^n$ and $\{p_k\}_{k=1}^n$ are encoded into the vector $s$.
The quantity $w(H,s)$ effectively counts the white vertices that possess an even number of black neighbours, with the external parity accounting for potential black neighbours external to the considered graph.
A transfer matrix tracks the weight changes between different states of two graphs.
In the ``cut-and-glue" approach, the transfer matrix can be expressed as a product of two simpler transfer matrices: one for each operation.
\Cref{fig:recursive_construction} shows an example of the ``cut-and-glue'' procedure to build the path, star and cycle graph families in a recursive way.

The first operation corresponds to an \emph{evolution} from the graph $H$ to the graph $J$.
It is represented by
\begin{equation}
    E^{J \leftarrow H} \in \mathbb{R}[y]^{4^{|J|} \times 4^{|H|}}.
\end{equation}
The entry $E^{J \leftarrow H}_{ij}$ encodes the weight update when $H$ is in state $j$ and $J$ in state $i$:
\begin{equation}
    E^{J \leftarrow H}_{ij} = 
    \begin{cases}
        x^{\Delta}y^{|J|-|H|-\Delta}, & i \text{ is compatible with } j, \\
        0, & \text{otherwise,}
    \end{cases}\nonumber
\end{equation}
where $\Delta := w(J,i)-w(H,j)$ and compatibility requires that for all $k \in V(H)$ we have that
\begin{equation}
\begin{aligned}
    c_k^{(i)} &= c_k^{(j)}, \\
    p_k^{(i)} &= p_k^{(j)}.
\end{aligned}
\end{equation}
Note that here we interpreted the $k$ vertices as vertices in $J$, since there is a one-to-one map from $H$ to $J$,

The second operation corresponds to the \emph{restriction} of the graph $J$ to a subgraph $H'$. 
It is represented by the transfer matrix
\begin{equation}
    R^{H' \leftarrow J} \in \mathbb{R}[y]^{4^{|H'|} \times 4^{|J|}}.
\end{equation}
The entry $R^{H' \leftarrow J}_{ij}$ encodes the weight update when $J$ is in state $j$ and $H$ in state $i$:
\begin{equation}
    R^{H' \leftarrow J}_{ij} = 
    \begin{cases}
        1, & i \text{ is compatible with } j, \\
        0, & \text{otherwise,}
    \end{cases}
\end{equation}
where compatibility requires that the colours and the parities of $J$ and $H'$ match. 
Note that for the parity of $H'$, we have to account for the vertices that were in $J$ and no longer appear in $H'$.
Formally, a state $i$ of $H'$ is compatible with a state $j$ of $J$, if for all vertices $k \in V(H)$:
\begin{equation}
\begin{aligned}
    c_k^{(i)} &= c_k^{(j)}, \\
    p_k^{(i)} &= p_k^{(j)} + \sum_{l \in N_{J \setminus H'}(k)} c_l^{(j)} \bmod 2,
\end{aligned}
\end{equation}
with $N_{J \setminus H'}(k)$ denoting the neighbours of $k$ that are not in $H'$, and $c_k^{(i)}, p_k^{(i)}, c_k^{(j)}, p_k^{(j)} \in \{0,1\}$ representing the colour and parity of node $k$ in states $i$ and $j$, respectively.

In the case of recursive graphs, the ``usual" transfer matrix corresponds to the evolution from the graph $H$ to the graph $J$ followed by a reduction to the subgraph $H'$:
\begin{equation}
    T^{H' \leftarrow H} := R^{H' \leftarrow J} \, E^{J \leftarrow H}  \in \mathbb{R}[y]^{4^{|H'|} \times 4^{|H|}}.
\end{equation}
As a consequence of this description, the initial vector corresponds to the evolution from the empty graph to the initial graph $G$ followed by a reduction to the subgraph $H$:
\begin{equation}
    \boldsymbol{v} := R^{H \leftarrow G} E^{G \leftarrow \emptyset},
\end{equation}
whose dimensions reduce to a column vector. 

The whole construction process of a recursively defined graph can be visualized as:
\begin{equation}
    \emptyset \underbrace{ \overset{E}{\longrightarrow} G \overset{R}{\longrightarrow}}_{\boldsymbol{v}} H \underbrace{\overset{E}{\longrightarrow} J \overset{R}{\longrightarrow}}_{T^{H' \leftarrow H}} H' \dots
\end{equation}
The last step is recursively repeated. 
This recursion is possible, as long as $H$ and $H'$ are isomorphic.

Finally, when building the generating function, one should not forget to multiply each transfer matrix $E^{J \leftarrow H}$ by $z$ to update the ``step" index.
Thus, the GF for recursively defined graphs is given by
\begin{equation}
\begin{aligned}
    \gf(x,y,z) & = 1 + z\left\| \sum_{r = 0}^{\infty}  z^r \left( T^{H' \leftarrow H} \right)^r \boldsymbol{v} \right\|_{+} \\
    & = 1 + z \left\| (\mathbb{I} - zT^{H' \leftarrow H})^{-1} \boldsymbol{v} \right\|_{+}\label{eq:gen_func_TM}
\end{aligned}
\end{equation}
where we took the convention that the first graph corresponds to the empty graph with WEP $W_0(x,y) = 1$.

Generating functions associated with recursively defined polynomial sequences are necessarily rational, that is, they can be expressed as a ratio of polynomials,
\begin{equation}
\gf(x,y,z) = \frac{p(x,y,z)}{q(x,y,z)}\ .
\end{equation}
This follows from the representation of the generating function in Eq.~\eqref{eq:gen_func_TM}, which is based on the inverse of the matrix $\mathbb{I}-zT^{H' \leftarrow H}$. 
Whenever this inverse exists, it can be written as the adjugate divided by the determinant of $\mathbb{I}-zT^{H' \leftarrow H}$. 
Since both the adjugate and the determinant are polynomial functions of $(x,y,z)$, the resulting generating function is rational.

We note that families of recursively defined graphs can be extended by artificially adding graphs at the beginning of the family. 
For example, in the cycle graph family, we include the single-vertex graph and the graph consisting of two disconnected vertices for convenience. 
However, these graphs do not correspond to proper cycle graphs, and their WEPs cannot be described using the transfer-matrix approach.
The generating function generalizes to 
\begin{equation}
\begin{aligned}
    \gf(x,y,z) = &~1 + \sum_{r=1}^R W_r(x,y) z^r \\
    & + z^{R+1} \left\| (\mathbb{I} - zT)^{-1} \boldsymbol{v} \right\|_{+}
\end{aligned}
\end{equation}
where the WEP of the first $R$ graphs are computed separately and $T$ is the transfer matrix.
We say in this case that the recursion starts at $R+1$. 

\section{Coefficients of a rational function} \label{app:asymptotic_behaviour_rational_functions}

\paragraph{Exact computation.}
Suppose we have a rational function
\begin{equation}
    f(z) = \frac{p(z)}{q(z)},
\end{equation}
where $p(z)$ and $q(z)$ are polynomials with $\textrm{deg}~p < \textrm{deg}~q$.  
We assume that $f$ is simplified, i.e.~$p$ and $q$ have no common factors. 
Moreover, suppose $f$ is analytic at $z=0$, i.e.~$q(0) \neq 0$, so that it 
can be written as a Taylor series around $z=0$,
\begin{equation}
    f(z) = \sum_{r=0}^{\infty} a_r z^r .
\end{equation}
The coefficients $a_r$ can be computed either by differentiation (Taylor's coefficient formula),
\begin{equation}
   a_r = \frac{f^{(r)}(0)}{r!},
\end{equation}
or by integration (Cauchy’s coefficient formula and the residue theorem),
\begin{equation}
    a_r
    = - \sum_{z_0 \in \{z : q(z) = 0\}}
      \operatorname{Res}_{z=z_0}
      \left( \frac{f(z)}{z^{r+1}} \right).
\end{equation}
In particular, if $z_0$ is a simple pole of $f(z)$, then
\begin{equation}
    \operatorname{Res}_{z=z_0}
    \left( \frac{f(z)}{z^{r+1}} \right)
    = \frac{p(z_0)}{q'(z_0)\, z_0^{\,r+1}} .
\end{equation}

\paragraph{Asymptotic approximation.} 
We define the \emph{dominant singularity} $z_*$ as the root of $q(z)$, which has the smallest modulus, i.e.,
\begin{equation} \label{eq:dominant_singularity}
    z_* = \arg \min_{\{z : q(z) = 0\}} |z| \, ,
\end{equation}
and denote its multiplicity by $m$.
In other words, $q(z)$ can be factorized by $(z-z_*)^m$.
Note that in general, the dominant singularity is not necessarily unique.
In this paper, the functions satisfy the property that the $a_r$ are real and non-negative; this follows from the fact that the generating functions count objects.
In such a case, Pringsheim's theorem ensures that the dominant singularity is real and non-negative~\cite{flajolet_analytic_2009}. 

When $f$ is rational and the dominant singularity is unique, the behaviour of the coefficient $a_r$ (as $r \rightarrow \infty$) follows,
\begin{equation}\label{eq:approx_coeff_rational_function}
    a_r \sim (-1)^m \frac{m p(z_*)}{\partial_z^m q(z_*)} r^{m-1} z_*^{-r-m} \, .
\end{equation}
This is a direct consequence of Cauchy's integral formula. 
In the case where the unique dominant singularity has multiplicity 1, the approximation in \Cref{eq:approx_coeff_rational_function} becomes a purely exponential function of $r$:
\begin{equation} \label{eq:exponential_behaviour}
    a_r \sim c z_*^{-r} \, ,
\end{equation}
where $c :=  - \frac{p(z_*)}{q'(z_*)} z_*^{-1}$ and $z_*$ is given in \Cref{eq:dominant_singularity}.

In all cases, we observed that the generating functions obtained in this work possess a unique dominant singularity of multiplicity one. 
Consequently, \Cref{eq:exponential_behaviour} applies. 
Note that the dominant singularity being unique and of multiplicity $1$ is not too surprising.  The dominant singularity not being unique would lead to a sum of terms of the form in~\eqref{eq:approx_coeff_rational_function}, but where the terms corresponding to the non-dominant singularities not on the positive real axis would lead to oscillatory behaviour. 
Furthermore, any $m>1$ in Eq.~\eqref{eq:approx_coeff_rational_function} would predict an asymptotic behaviour of the fidelity (see the main text) that is not purely exponential. 
We would not expect such behaviour in a sequence of recursively definable graphs however, due to the limited `interactions' between $\ket{\Gamma_r}$ and $\ket{\Gamma_{r+1}}$.

\section{Combinatorial proofs of the generating functions}\label{app:comb_proofs_GF}
One of the powerful features of analytic combinatorics is that dense combinatorial constructions can be efficiently expressed in algebraic terms. 
As such, one can give more direct combinatorial proofs for the GFs of certain simple sequences of graph states. 
We give such proofs here now for star graphs, path graphs and cycle graphs. 
The proof for the cycle graph is quite similar in setup to the proof of~\cite{miller_shor-laflamme_2023} of the WEP for a single instance of an $n$-cycle graph. 
However, since generating functions are in general nicer to handle than the individual WEPs, our proof is significantly simpler.

We will make use of standard tools in analytic combinatorics~\cite{flajolet_analytic_2009}. 
For example, assume we have two sets of objects $A$ and $B$, where every object $a\in A$ has, say, parameters $n=n(a)$ and $k=k(a)$ with integer values, and the same for all $b\in B$. 
We now attach the following generating functions to $A$ and $B$, respectively: $\mathcal{A}=\sum_{a\in A}x^{n(a)}y^{k(a)}$ and $\mathcal{B}=\sum_{b\in B}x^{n(b)}y^{k(b)}$. 
Note that the product $\mathcal{A}\times \mathcal{B} := \mathcal{C}$ has an expansion $\mathcal{C} = \sum_{n, k} c_{n, k}~x^ny^k$, where $c_{n, k}$ counts exactly the number of ways to pick some $a\in A$ and $b\in B$ such that $n(a)+n(b) = n$ and $y(a)+y(b) = y(c)$. 
We thus see that the above combinatorial problem of picking objects from $A$ and $B$ with certain constraints can be conveniently kept track of through purely algebraic expressions.

\subsection{Combinatorial description of the star graph GF}
Up to local rotations, the stabilizers of the star graph are those of the GHZ state, i.e.~products of the all $X$-string, and even weight Pauli strings containing only $I$ and $Z$. 
Split these stabilizers into two types: those that are even weight Pauli strings $I,Z$ strings, and those that are not. 
Note that those stabilizers that do not consist of $I$ and $Z$ are full-weight.

First, notice that the following GF naturally counts the ways of picking $k$ objects out of $n$ objects,

\begin{align}
&~\frac{1}{1-(x+y)z}\\
= &~1 + (x+y)z + (x+y)^2z^2 + (x+y)^3z^3 + \cdots\nonumber \\
= &~\sum_{n=0}^{\infty}\left(\sum_{k=0}^n \binom{n}{k}x^{n-k}y^{k} \right)z^n\\
\equiv&~E(x, y, z) ,\nonumber
\end{align}
where we used the binomial theorem. 
From this, a standard roots-of-unity filter shows that the generating function which counts only selections of an \emph{even} number $k$ of objects is $\frac{1}{2}\left(E(x, y, z) + E(x, -y, z)\right)$, corresponding to the even-weight stabilizers consisting of $I$ and $Z$. 
This is precisely the second term in the GF of the star graphs.

It remains to count the remaining $2^{n-1}$ stabilizers of weight $n$, 
\begin{align}
\frac{yz}{1-2yz} = &~0 + yz + 2y^2z^2 + 4y^3z^3 + \cdots \nonumber \\
=&\sum_{n=1}^{\infty} 2^{n-1}y^{n} z^n\ , \nonumber
\end{align}
which we see corresponds to the first term of the GF.

\subsubsection{Combinatorial description of the path graph GF}
We show here now how to count the number of colourings of a given weight for path graphs. 
Before proceeding, we will restrict ourselves to calculating $\mathcal{W}(x, 1, z)$ (i.e.~setting $y=1$). 
This simplifies notation, yet at the same time $\mathcal{W}(x, y, z)$ can be reconstructed from $\mathcal{W}(x, 1, z)$, since $\mathcal{W}(x, y, z)=\mathcal{W}(x/y, 1, yz)$.

First, let us consider the contribution of the all-white colourings, labeled by $\emptyset, \white, \white\white, \white \white\white, \ldots, $. 
The associated GF is given by

\begin{align} \label{eq:path_GF1}
\underbrace{\emptyset}_{1} + \underbrace{\white}_{xz} + \underbrace{\white \white}_{x^2z^2} + \underbrace{\white \white \white}_{x^3z^3} + \cdots 
=  \frac{1}{1-xz} . 
\end{align}

We can now consider all colourings with at least one black vertex. 
Note that such colourings can be decomposed in three parts. 
First, a part consisting of an arbitrary number of white vertices (including $0$ white vertices), followed by a black vertex. 
These contribute

\begin{equation}  \label{eq:path_GF2}
\begin{aligned}
\underbrace{\black}_{z} + \underbrace{\white \black}_{z^2}+\underbrace{\white\white \black}_{xz^3} + \underbrace{\white\white\white \black}_{x^2z^4} + \cdots \\
=z+\frac{z^2}{1-xz}\ .
\end{aligned}
\end{equation}

The second part consists of any number of similar `pieces' of the form $\black, \white \black, \white \white \black, \white \white \white \black, \ldots$. 
Note that, unlike the first part, these piececs will contribute different weights, since they will all have a black vertex preceding them. 
Adding a single such piece thus contributes

\begin{equation}
    \begin{aligned}
     \underbrace{\black}_{z} + \underbrace{\white \black}_{xz^2} + \underbrace{\white\white \black}_{z^3} + \underbrace{\white\white\white \black}_{xz^4} + \cdots \\
        \rightarrow z+xz^2 + \frac{z^3}{1-xz} \ . 
    \end{aligned}
\end{equation}

Since we can have any number $k$ of such pieces, we find a total contribution of 

\begin{equation} \label{eq:path_GF3}
\begin{aligned}
\sum_{k=0}^\infty \left(z + xz^2 + \frac{z^3}{1-xz}\right)^k = \frac{1}{1-\left(z + xz^2 + \frac{z^3}{1-xz}\right)}\ . 
\end{aligned}
\end{equation}

Finally, the all-white pieces at the end contribute a factor of $1+\frac{z}{1-xz}$, since 

\begin{equation} \label{eq:path_GF4}
\begin{aligned}
\underbrace{\emptyset}_{1}+\underbrace{\white}_{z}+ \underbrace{\white \white}_{xz^2}+ \underbrace{\white \white \white}_{x^2z^3}+\underbrace{\white \white \white \white}_{x^3z^4} \ldots\\
\rightarrow 1+\frac{z}{1-xz} \ .
\end{aligned}
\end{equation}

Now since the set of colourings is equal to the disjoint union of the colourings with only white vertices (Eq.~\ref{eq:path_GF1}), and the Cartesian product of the pieces described in Eqs.~\ref{eq:path_GF2},~\ref{eq:path_GF3} and \ref{eq:path_GF4}, we find that the generating function for the path graph is given by

\begin{equation}
\begin{aligned}
\frac{1}{1-xz} + \left(z+\frac{z^2}{1-xz}\right)\\
\times \left(\frac{1}{1-\left(z + xz^2 + \frac{z^3}{1-xz}\right)}\right) \left(1+\frac{z}{1-xz}\right)\\
= \frac{1-2(x-1)z^2}{1-\left(\left(1+x\right)z-\left(1-x^2\right)z^3\right)} ,
\end{aligned}
\end{equation}
which we see indeed equals $\mathcal{W}(x, 1, z)$ for the family of path graphs.

\subsection{Combinatorial description of the cycle graph GF}
The enumeration of the colourings of the cycle graph family can be divided into three cases.
For the first case, we consider the contributions when all the vertices are white, from which we find from \Cref{eq:path_GF1} that it is given by $\frac{1}{1-xz}$.

For the second case, we look at the contribution from colourings with exactly one black vertex which is not the first one:
\begin{equation}
\begin{aligned}
    \underbrace{\white \black}_{x z^2} 
    + 
    \underbrace{
        \begin{aligned}
            \white \black \white \\
            \white \white \black
        \end{aligned}
    }_{2 z^3}
    + 
    \underbrace{
        \begin{aligned}
            \white \black \white \white \\
            \white \white \black \white \\
            \white \white \white \black \\
        \end{aligned}
    }_{3xz^4} 
    + \dots \\
    \rightarrow x z^2 + z^3\sum_{k=0}^{\infty} (2+k) x^k z^k \\
    = xz^2 + 2z^2 \sum_{k=0}^{\infty}(xz)^k + z^4 \frac{\partial}{\partial z} \left( \sum_{k=0}^{\infty} (xz)^k \right) \\
    = xz^2 + \frac{2z^3}{1-xz} + \frac{xz^4}{(1-xz)^2}\ .
\end{aligned}
\end{equation}
Note that we assumed that the 2-vertex graph is disconnected.

\begin{figure}[ht!]
    \centering
    \includegraphics[width=0.7\linewidth]{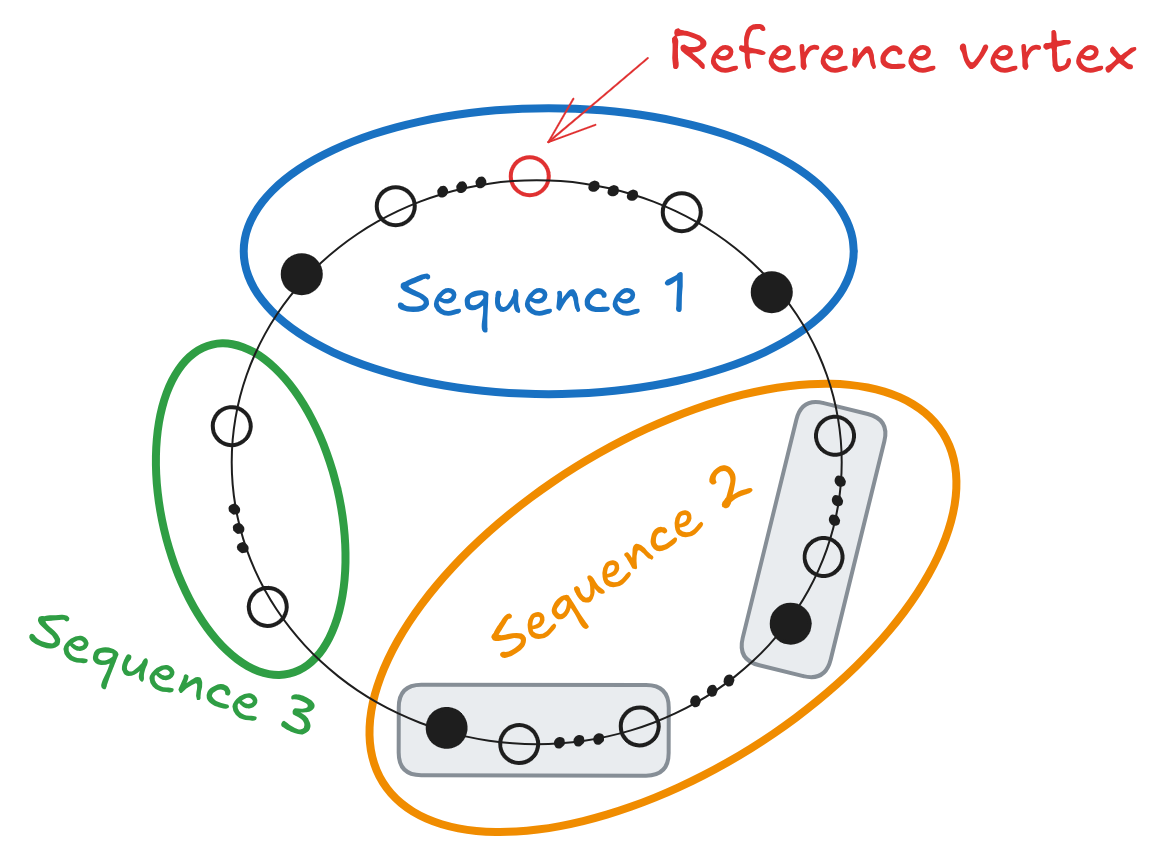}
    \caption{Each colouring in the third case can be expressed as three sequences of vertices.
    To distinguish colourings, we fix a reference vertex.}
    \label{fig:combinatorics_cycle_graph}
\end{figure}

The third case corresponds to the remaining colourings.
That is, either the first vertex is black, or the first vertex is white and there are at least two black vertices.
This last case can be visualized in \Cref{fig:combinatorics_cycle_graph}.

The first sequence goes from the first black vertex on the left of the reference vertex to the first black vertex on the right.
\begin{equation}
\begin{aligned}
    \underbrace{\black}_{z} 
    + 
    \underbrace{\black {\color{red}\fullmoon} \black}_{xz^3}
    + 
    \underbrace{
    \begin{aligned}
        \black {\color{red}\fullmoon} \white \black \\
        \black \white {\color{red}\fullmoon} \black
    \end{aligned}}_{2z^4}
    + 
    \underbrace{
    \begin{aligned}
        \black {\color{red}\fullmoon} \white \white \black \\
        \black \white {\color{red}\fullmoon} \white \black \\
        \black \white \white {\color{red}\fullmoon} \black
    \end{aligned}}_{3xz^5} 
    + \dots \\
    \rightarrow z + xz^3 +  z^4 \sum_{k=0}^{\infty} (2+k) x^k z^k \\
    = z + xz^3 + \frac{2 z^4}{1-x z} + \frac{x z^5}{(1-x z)^2}
\end{aligned}
\end{equation}
where the red vertex corresponds to the reference vertex.

The second sequence exactly corresponds to \Cref{eq:path_GF3}.
As with the path graph GF, the third sequence corresponds to a sequence of white vertices. 
We have that both the left- and right-most vertices are both adjacent to a black vertex, as opposed to the path graph GF where only the left-most vertex was adjacent to a black vertex.
\begin{equation}
    \begin{aligned}
        \underbrace{\emptyset}_{1} 
        +
        \underbrace{\white}_{xz}
        +
        \underbrace{\white \white}_{z^2}
        +
        \underbrace{\white \white \white}_{x z^3} 
        + \dots\\
        \rightarrow 1 + xz + z^2 \sum_{k=0}^{\infty} x^kz^k \\
        = 1 + xz + \frac{z^2}{1-xz}
    \end{aligned}
\end{equation}

Finally combining all these generating functions together, we find
\begin{equation}
    \begin{aligned}
        \frac{1}{1-xz} + xz^2 + \frac{2z^3}{1-xz} + \frac{xz^4}{(1-xz)^2} \\
        + \left( z + xz^3 + \frac{2 z^4}{1-x z} + \frac{x z^5}{(1-x z)^2} \right) \\
        \times \left( \frac{1}{1-(z + xz^2 + \frac{z^3}{1-xz})} \right) \left( 1 + xz + \frac{z^2}{1-xz} \right) \\
        = \frac{1-2 \left(x-1\right) \left(x+1\right) z^3}{1 -z(x+1)( 1+ \left(x-1\right) z^2)}
    \end{aligned}
\end{equation}
which we see indeed equals $\mathcal{W}(x, 1, z)$ for the family of cycle graphs.
To retrieve the expression from \Cref{eq:gf_cycle_graph_family}, we have to apply the substitutions $x \mapsto x/y$ and $z \mapsto yz$.

\section{More examples} \label{app:more_examples}

We give the numerator and denominator of various generating functions.

\paragraph{Complete bipartite graph}
\begin{equation}
\begin{aligned}
    p(x,y,z) = & -2x^3 y^2 z^5 + 4 x^3 y z^4 - x^3 z^3 + 2 x^2 y^3 z^5 \\
    & + 4 x^2 y^2 z^4 - 8 x^2 y z^3 + 2 x^2 z^2 + 2 x y^4 z^5 \\
    & - 4 x y^3 z^4 - 3 x y^2 z^3 + 6 x y z^2 \\
    & + x z (-2 x^2 y z^3 + x^2 z^2 + 4 x y z^2 - 2 x z \\
    & + 2 y^3 z^3 - y^2 z^2 - 2 y z + 1) \\
    & - 2 x z - 2 y^5 z^5 - 4 y^4 z^4 + 4 y^3 z^3 \\
    & + y z (-2 x^2 y z^3 + x^2 z^2 + 4 x y z^2 - 2 x z \\
    & + 2 y^3 z^3 - y^2 z^2 - 2 y z + 1) - 2 y z + 1 \\
   q(x,y,z) = & -2 x^2 y z^3 + x^2 z^2 + 4 x y z^2 - 2 x z + 2 y^3 z^3 \\
   & - y^2 z^2 - 2 y z + 1
\end{aligned}
\end{equation}

\paragraph{Pusteblume graph}
\begin{equation}
    \begin{aligned}
        p(x,y,z) = & 2 x^5 y z^3 - x^5 z^2 + 3 x^4 y^2 z^3 - 2 x^4 y z^2\\
        & + x^4 z + 4 x^3 y^3 z^3 - 8 x^3 y^2 z^2 + 2 x^2 y^4 z^3 \\
        & - 6 x^2 y^3 z^2 + 6 x^2 y^2 z - 2 x^2 y z^3 + x^2 z^2 \\
        & - 6 x y^5 z^3 - 7 x y^4 z^2 + 4 x y z^2 - 2 x z - 5 y^6 z^3 \\
        & - 8 y^5 z^2 + 9 y^4 z + 2 y^3 z^3 - y^2 z^2 - 2 y z + 1\\
        q(x,y,z) = & -2 x^2 y z^3 + x^2 z^2 + 4 x y z^2 - 2 x z + 2 y^3 z^3 \\
        & - y^2 z^2 - 2 y z + 1
    \end{aligned}
\end{equation}

\paragraph{Joint squares graph}
\begin{equation}
    \begin{aligned}
        p(x,y,z) = & x^6 y z^2 + 3 x^5 y^2 z^2 - x^5 y z^2 \\
        & + 4 x^4 y^3 z^2 - 2 x^4 y^2 z^2 - x^4 z + 2 x^3 y^4 z^2\\
        & - 2 x^3 y^3 z^2 + x^3 z - 3 x^2 y^5 z^2 - 2 x^2 y^2 z \\
        & + x^2 y z - 5 x y^6 z^2 + 3 x y^5 z^2 - 8 x y^3 z + 3 x y^2 z \\
        & - 2 y^7 z^2 + 2 y^6 z^2 - 5 y^4 z + 3 y^3 z - 1 \\
        q(x,y,z) = & -x^5 y z^2 - 2 x^4 y^2 z^2 - 2 x^3 y^3 z^2 + x^3 z + x^2 y z \\
        & + 3 x y^5 z^2 + 3 x y^2 z + 2 y^6 z^2 + 3 y^3 z - 1
    \end{aligned}
\end{equation}

\paragraph{$n \times 2$ grid graph}
\begin{equation}
\begin{aligned}
p(x,y,z) = & -4 x^6 y^4 z^5 + 8 x^5 y^5 z^5 + 4 x^4 y^6 z^5 + 3 x^4 y^2 z^3 \\
& - 16 x^3 y^7 z^5 + 4 x^2 y^8 z^5 - 6 x^2 y^4 z^3 + 4 x^2 y^2 z^2 + 8 x y^9 z^5 \\
& - 8 x y^3 z^2 - 4 y^10 z^5 + 3 y^6 z^3 + 4 y^4 z^2 - 1 \\
q(x,y,z) = &x^8 y^4 z^6 - 4 x^6 y^6 z^6 - x^6 y^2 z^4 + 6 x^4 y^8 z^6 - x^4 y^4 z^4 \\
& - 2 x^4 y^2 z^3 - 4 x^2 y^10 z^6 + 5 x^2 y^6 z^4 + 4 x^2 y^4 z^3 \\
& + x^2 z + y^12 z^6 - 3 y^8 z^4 - 2 y^6 z^3 + 3 y^2 z - 1
\end{aligned}
\end{equation}

\end{document}